\documentclass[fleqn,usenatbib]{mnras}
\usepackage{amsmath}
\usepackage{xspace}
\usepackage{mathptmx}
\usepackage{txfonts}
\usepackage[dvipsnames]{xcolor}

\usepackage[T1]{fontenc}
\usepackage{ae,aecompl}

\usepackage{graphicx}	
\usepackage{amssymb}	
\usepackage{booktabs}
\usepackage{wasysym}
\usepackage[normalem]{ulem}
\usepackage{pdflscape}
\usepackage{soul}

\let\oldhref\href
\renewcommand{\href}[2]{\oldhref{#1}{\hbox{#2}}}

\definecolor{colorl1}{RGB}{0, 51, 153}
\definecolor{colorl2}{RGB}{153, 0, 0}
\definecolor{colorl3}{RGB}{179, 179, 0}
\definecolor{colorl4}{RGB}{51, 102, 0}

\definecolor{colorw1}{RGB}{51, 102, 255}
\definecolor{colorw2}{RGB}{255, 51, 0}
\definecolor{colorw3}{RGB}{255, 214, 51}
\definecolor{colorw4}{RGB}{51, 204, 51}
\definecolor{seagreen}{rgb}{0.190, 0.525, 0.361} 

\newcommand{\ahf}{\texttt{AHF}}

\newcommand{\hMpc}{{\ifmmode{h^{-1}{\mathrm Mpc}}\else{$h^{-1}$Mpc}\fi}}
\newcommand{\Mpc}{{\ifmmode{{\mathrm Mpc}}\else{Mpc}\fi}}
\newcommand{\hkpc}{{\ifmmode{h^{-1}{\mathrm kpc}}\else{$h^{-1}$kpc}\fi}}
\newcommand{\kpc}{{\ifmmode{ {\mathrm kpc}}\else{{\mathrm kpc}}\fi}}
\newcommand{\kms}{{\ifmmode{ {\mathrm km\,s^{-1}}}\else{ ${\mathrm km\,s^{-1}}$}\fi}}
\newcommand{\hMsun}{{\ifmmode{h^{-1}{\mathrm {M_{\astrosun}}}}\else{$h^{-1}{\mathrm{M_{\astrosun}}}$}\fi}}
\newcommand{\Msun}{{\ifmmode{{\mathrm M}_{\astrosun}}\else{${\mathrm M}_{\astrosun}$}\fi}}
\newcommand{\Mhalo}{{\ifmmode{M_{\mathrm halo}}\else{$M_{\mathrm halo}$}\fi}}
\newcommand{\Rvir}{{\ifmmode{R_{\mathrm{vir}}}\else{$R_{\mathrm{vir}}$}\fi}}
\newcommand{\Rgal}{{\ifmmode{R_{\mathrm{gal}}}\else{$R_{\mathrm{gal}}$}\fi}}
\newcommand{\Rtwohun}{{\ifmmode{R_{200}}\else{$R_{200}$}\fi}}
\newcommand{\Mvir}{{\ifmmode{M_{\mathrm{vir}}}\else{$M_{\mathrm{vir}}$}\fi}}
\newcommand{\Mtwohun}{{\ifmmode{M_{200}}\else{$M_{200}$}\fi}}
\newcommand{\Nvir}{{\ifmmode{N_{\mathrm vir}}\else{$N_{\mathrm vir}$}\fi}}
\newcommand{\Tvir}{{\ifmmode{T_{\mathrm{vir}}}\else{$T_{\mathrm{vir}}$}\fi}}
\newcommand{\Tcutoff}{{\ifmmode{T_{\mathrm{cutoff}}}\else{$T_{\mathrm{cutoff}}$}\fi}}
\newcommand{\Tmax}{{\ifmmode{T_{\mathrm{max}}}\else{$T_{\mathrm{max}}$}\fi}}
\newcommand{\Mstar}{{\ifmmode{M_\star}\else{$M_\star$}\fi}}
\newcommand{\Mshock}{{\ifmmode{M_{\mathrm{shock}}}\else{$M_{\mathrm{shock}}$}\fi}}
\newcommand{\Mstream}{{\ifmmode{M_{\mathrm{stream}}}\else{$M_{\mathrm{stream}}$}\fi}}
\newcommand{\Vrot}{{\ifmmode{V_{\mathrm rot}}\else{$V_{\mathrm rot}$}\fi}}
\newcommand{\Reff}{{\ifmmode{R_{\mathrm e}}\else{$R_{\mathrm e}$}\fi}}
\newcommand{\Sigmae}{{\ifmmode{\Sigma_{\mathrm e}}\else{$\Sigma_{\mathrm e}$}\fi}}
\newcommand{\logmstar}{{\ifmmode \log(M_\star/\mathrm{M}_\odot) \else $\log(M_\star/\mathrm{M}_\odot)$ \fi}}
\newcommand{\ltsima}{$\; \buildrel < \over \sim \;$}
\newcommand{\gtsima}{$\; \buildrel > \over \sim \;$}
\newcommand{\lsim}{\lower.5ex\hbox{\ltsima}}
\newcommand{\gsim}{\lower.5ex\hbox{\gtsima}}

\def\lesssim{\mathrel{\hbox{\rlap{\hbox{\lower4pt\hbox{$\sim$}}}\hbox{$<$}}}}
\def\gtrsim{\mathrel{\hbox{\rlap{\hbox{\lower4pt\hbox{$\sim$}}}\hbox{$>$}}}}

\newcommand{\beq}{\begin{equation}}
\newcommand{\eeq}{\end{equation}}
\def\beqa{\begin{eqnarray}}
\def\eeqa{\end{eqnarray}}
\def\LCDM{\ensuremath{\Lambda}CDM}

\def\head{ \vbox to 0pt{\vss \hbox to 0pt{\hskip 440pt\mathrm
      LA-UR-10-07069\hss} \vskip 25pt}}

\def \kms {\ifmmode\,\mathrm km\,s^{-1}\else$\,\mathrm km\,s^{-1}$\fi}
\def \kpc {\ifmmode{\,\mathrm kpc}\else${\mathrm  kpc}$\fi}  
\def \hkpc {\ifmmode{h^{-1}\mathrm kpc}\else${h^{-1}\mathrm kpc}$\fi}  
\def \hMpc {\ifmmode{h^{-1}\mathrm Mpc}\else${h^{-1}\mathrm Mpc}$\fi}  
\def \Mpch {\ifmmode{h^{-1}\mathrm Mpc}\else${h^{-1}\mathrm Mpc}$\fi}  
\def \Msun {\ifmmode{\mathrm M}_{\astrosun}\else${\mathrm M}_{\astrosun}$\fi}
\def \hMsun {\ifmmode h^{-1}\,\mathrm M_{\astrosun}\else$h^{-1}\,\mathrm M_{\astrosun}$\fi}
\def \Gyr {\ifmmode\,\mathrm Gyr\else$\,$Gyr\fi}

\def \LCDM {\ifmmode\Lambda{\mathrm CDM}\else$\Lambda{\mathrm CDM}$\fi}
\def \sig8 {\ifmmode\sigma_8\else$\sigma_8$\fi} 
\def \OmegaM {\ifmmode\Omega_{\mathrm{m}}\else$\Omega_{\mathrm{m}}$\fi} 
\def \Omegab {\ifmmode\Omega_{\mathrm{b}}\else$\Omega_{\mathrm{b}}$\fi} 
\def \OmegaL {\ifmmode\Omega_{\mathrm\Lambda}\else$\Omega_{\mathrm\Lambda}$\fi} 
\def \Deltavir {\ifmmode\Delta_{\mathrm vir}\else$\Delta_{\mathrm vir}$\fi}
\def \rhocrit {\ifmmode\rho_{\mathrm crit}\else$\rho_{\mathrm crit}$\fi}
\def \rhou {\ifmmode\rho_{\mathrm u}\else $\rho_{\mathrm u}$\fi}
\def \zc {\ifmmode z_{\mathrm c}\else $z_{\mathrm c}$\fi}

\def \fcold {\ifmmode{f_\mathrm{cold}}\else${f_\mathrm{cold}}$\fi\xspace}

\newcommand{\Params}{\boldsymbol{\Theta}\xspace}
\newcommand{\Hyper}{\boldsymbol{\mathcal{H}}\xspace}
\newcommand{\Data} {\boldsymbol{\mathcal{D}}\xspace}

\newcommand{\Prob}{\mathrm{P}}

\newcommand{\diff}{\mathrm{d}}

\title[Gas accretion at high $z$] {Gas accretion at high redshift: cold flows all the way}

\author[
S.\ Waterval, C.\ Cannarozzo and A.\ V.\ Macciò]{Stefan Waterval$^{\href{https://orcid.org/0000-0002-5542-8624}{\hskip2pt\includegraphics[width=9pt]{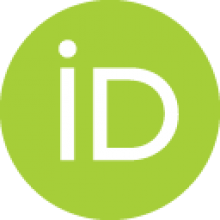}}{1,2}}$\thanks{E-mail: sw4445@nyu.edu},
Carlo Cannarozzo$^{\href{https://orcid.org/0000-0003-3843-7366}{\hskip2pt\includegraphics[width=9pt]{figures/orcid-ID.png}}{1,2}}$,
and Andrea V.\ Macciò$^{\href{https://orcid.org/0000-0002-8171-6507}{\hskip2pt\includegraphics[width=9pt]{figures/orcid-ID.png}}{1,2,3}}$
\\
$^{1}$New York University Abu Dhabi, PO Box 129188, Abu Dhabi, United Arab Emirates \\
$^2$Center for Astrophysics and Space Science (CASS), New York University Abu Dhabi\\  
$^3$Max-Planck-Institut f\"ur Astronomie, K\"onigstuhl 17, D-69117 Heidelberg, Germany
}

\date{Accepted 2025 January. Received 2024 October; in original form 2024 September}

\pubyear{2025}

\begin{document}
\label{firstpage}
\pagerange{\pageref{firstpage}--\pageref{lastpage}}
\maketitle

\begin{abstract}


We study in detail how massive galaxies accreate gas through cosmic time using cosmological hydrodynamical simulations from the High-$z$ Evolution of Large and Luminous Objects (HELLO) and the Numerical Investigation of a Hundred Astrophysical Objects (NIHAO) projects.
We find that accretion through cold filaments at high $z$ ($z\approx2$--4) is a key factor in maintaining the high star-formation rates ($\gtrsim 100\,\Msun\,\mathrm{yr}^{-1}$) observed in these galaxies, and that more than 75\% of the total gas participating in the star formation process is accreted via this channel at high $z$ even in haloes well above $10^{12}$ \Msun. 
The low volume occupancy of the filaments allows plenty of space for massive gas outflows generated by the vigorous star formation and AGN activity, with the cold incoming gas and the hot outflowing gas barely interacting.
We present a model based on Bayesian hierarchical formalism able to accurately describe the evolution of the cold fraction accretion with redshift and halo mass. Our model predicts a relatively constant critical mass ($M_{\mathrm{c}}$) for cold-to-hot transition up to $z\sim1.3$ and an evolving critical mass $\log(M_{\mathrm{c}}) \propto \log(1+z)^{1.7}$ at higher redshift.
Overall, our findings provide deeper insight into the cosmic evolution of gas accretion modes and offer a robust framework for understanding how cold accretion contributes to galaxy growth across different epochs.

\end{abstract}

\begin{keywords}
quasars: supermassive black holes, galaxies: formation, galaxies: evolution, methods: numerical, methods: statistical
\end{keywords}

\section{Introduction}\label{sec:introduction}




How galaxies form stars and grow involves a mixture of complex processes at different scales.
The root element is the accretion of gas that has to make its way from large distances to eventually form stars inside the galaxy.
Along the way, the gas must pass through the halo, where it interacts with the circumgalactic medium (CGM), undergoing processes such as cooling, shock heating, and potential feedback from supernovae (SNe) or active galactic nuclei (AGNs).
These interactions can significantly influence whether and how the gas eventually condenses to fuel star formation.
Understanding these complex, multi-scale processes is crucial for unraveling the connection between gas accretion, star formation, and the overall evolution of galaxies.
For extended reviews on the CGM focusing either on observations and simulations, see e.g. \citet{Tumlinson2017} and \citet{FaucherGiguere2023}.

The classical view of galaxy formation from pioneering works \citep[][]{Binney1977,Rees1977,Silk1977,White1978,White1991} theorises that as the cold and diffuse gas from the intergalactic medium (IGM) falls into the gravitational potential of a dark matter (DM) halo, it should encounter a shock, heating it to the halo virial temperature and leaving it in hydrostatic equilibrium where its gravitational pull is counterbalanced by its pressure support.

In order to turn into stars, the virialised gas has to condense further towards higher densities by cooling. The ability of the gas to cool efficiently depends upon three important timescales: the cooling time $t_{\rm{cool}}$, the dynamical timescale $t_{\rm{dyn}}$, and the Hubble time $t_H$.
It follows three different possible regimes conditioned on the relative values of these three timescales. In the first regime, where $t_H < t_{\rm{cool}}$, the cooling time is larger than the age of the Universe, making it impossible for the gas to lose its pressure support and remain in hydrostatic equilibrium. If the cooling time is shorter than the Hubble time but larger than the dynamical time, $t_{\rm{dyn}} < t_{\rm{cool}} < t_H$, the gas cools inefficiently. It slowly contracts but the system has enough time to respond against going out of equilibrium. It results in the gas being in quasi-static equilibrium, slowly making its way towards the center of the potential in a quasi-adiabatic fashion. The last and most efficient regime for star formation is when $t_{\rm{cool}} < t_{\rm{dyn}}$, i.e. the gas can rapidly radiate its energy away and with the ensuing loss of thermal pressure support, the gas accretion rate onto the galaxy is then set by the dynamical timescale.

The existence of stable shock in the outskirts of a DM halo is however not trivial. \citet{Birnboim2003} investigated the conditions required for a stable shock to expand and persist far from the halo center and found that there is a characteristic halo mass $\sim 10^{11}\,\Msun$ below which there is no shock. The gas thus remains cold throughout its entire journey inside the circumgalactic medium (CGM). These findings were soon after confirmed by \citet{Keres2005} who tracked the temperature history of accreted gas using smoothed particle hydrodynamics (SPH) simulations. They propose an empirical temperature threshold in the maximum temperature a gas particle ever reached, $T_{\rm{max}} = 2.5 \times 10^5\,\mathrm{K}$, to distinguish between two accretion modes: the \emph{cold mode} and the \emph{hot mode} \citep[see also,][]{Katz2003}{}{}. As the nomenclature suggests, the former consists of gas whose temperature never exceeded $10^5\,\mathrm{K}$, while the latter encompasses gas that exceeded that temperature and is believed to have been shock heated when entering the halo. Their findings reveal that the cold mode dominates for systems with halo mass $\lesssim 10^{11.4}\,\Msun$.

While there is a critical halo mass around $10^{12}\,\Msun$ above which gas particles are shock heated and where the hot mode increasingly becomes the prominent mode of gas accretion onto galaxies, there can still be collimated filaments of cold and dense gas able to pierce through the halo unshocked \citep[][]{Keres2005,Keres2009a,Dekel2006,Dekel2009a,Brooks2009,Voort2011a, FaucherGiguere2011,Voort2012, Dubois2012, Aung2024}, however see \citet{Nelson2013,Nelson2016} for a more nuanced interpretation. Moreover, these cold streams appear very stable against various perturbations \citep[][]{Wang2014} and are preserved by galactic winds escaping into voids \citep[][]{Theuns2002}.

The combination of cold mode accretion within a hot halo is believed to take place mostly at high-$z$, due to the fact that massive halos $\gsim 10^{12}$ at $z \geq 2$ tend to be cosmological nodes at the intersection of multiple thinner DM filaments whose cross-section is significantly smaller than halo cross-section. The cooling time of the gas embedded in these filaments is short due to the local higher density, allowing the fuel for star formation to be efficiently funnelled towards the galaxy while remaining cold. In their work, \citet{Dekel2006} model a redshift dependent characteristic mass allowing for `cold in hot' accretion. Their model (shown in their Fig.7) suggests that with increasing redshift, haloes of increasing mass above the cold-to-hot threshold can still sustain cold accretion. This implies a critical mass evolving in two regimes: below some transition redshift the critical mass is constant around $\sim\!10^{12}\,\Msun$, while above it the critical mass increases with redshift. In the rest of the paper, we refer to the critical mass in these two regimes as \Mshock\ and \Mstream, respectively.

In observations, detection of cold streams is of wide interest but difficult, due to their small cross-section. This problem gets amplified in massive haloes where the cold streams are embedded in the hot virialised gas, dominating emissions in X-rays. In spite of these challenges, numerous observations report detecting a significant amount of cold gas inside the CGM whose characteristics are consistent with cold flows \citep[][]{Maccio2006, Bouche2016, Prochaska2014, Borisova2016, Leclercq2017, ArrigoniBattaia2018, Martin2019}. More recently, \citet{Daddi2022a} report evidence in $2 < z < 3.3$ galaxies for a transition from  `cold in hot' to hot-dominated accretion consistent with theoretical predictions for \Mstream. Furthermore, \citet{Daddi2022b} study the evolution of the stellar mass at which the star-formation main sequence flattens \citep[see, e.g.][]{Schreiber2015, Tomczak2016,Popesso2023} and find the corresponding halo mass to be in excellent agreement with the predicted evolution of \Mstream.

So far, our attention has been focused on the `positive feedback' for star formation, i.e. the accretion of gas onto the galaxy. During its migration from multi-kpc scales towards the center of the halo, the accreting material can face multiple instances of `negative feedback' in the form of galactic winds originating from supernovae (SN) and active galactic nucleus (AGN) driven outflows \citep[][]{Chevalier1985,Murray2005,FaucherGiguere2012}{}{}, radiation pressure from the ultraviolet background \citep[UVB;][]{Efstathiou1992,Haardt1996,Haardt2001,FaucherGiguere:2009, Haardt2012}{}{}, local sources \citep[local photoionisation feedback, hereafter LPF;][]{Cantalupo2010,Gnedin2012,Kannan2014,Obreja2019}{}{}, and young massive stars.

Many groups have investigated the potential impact of several types of feedback processes on the rates, modes, and types of gas accretion. Accretion onto galaxies at $z \leq 1$ is dominated by recycled gas that was first accreted, then ejected by stellar feedback, and accreted again in possible multiple cycles \citep[][]{Oppenheimer2010, Woods2014, Ubler2014, Tollet2019, Tollet2022}. \citet{FaucherGiguere2011} find that net gas accretions rates onto the halo can be altered by galactic outflows. Contrasted results are found by \citet{Voort2011a}, where accretion onto the halo is relatively insensitive to feedback, however, accretion onto the galaxy is sensitive to radiative cooling and to SN and AGN-driven outflows. In a subsequent paper, \citet{Voort2011b} find AGN feedback to be paramount in hindering gas in the hot halo from accreting further to the galaxy. Similarly, \citet{Nelson2015} demonstrate that the smooth inflow of material contributing to the growth of galaxy is significantly reduced by stellar and AGN feedback at all redshifts. More recently, results from \citet{Correa2018a} show that SN feedback increases the mass of hot gas in the halo, in contrast to AGN feedback which expels it outside the halo, leading to a reduction of the hot gas mass. Furthermore, \citet{Correa2018b} notice that AGN feedback flattens the accretion rate onto galaxies for halo masses $\sim\!10^{12}\,\Msun$ at $z \leq 2$. Finally, results from \citet{Wright2020} confirm that baryonic feedback processes remove gas from haloes and at the same time suppress inflows onto them.

Combining simulations from the \textit{Numerical Investigation of a Hundred Astrophysical Objects} \citep[NIHAO;][]{Wang2015} and the \textit{High-$z$ Evolution of Large and Luminous Objects} (HELLO) recently introduced in \citet{Waterval2024}, we investigate the evolution of the cold fraction \fcold\ of gas accreted onto galaxies and their haloes and show that it decreases with mass and increases with redshift, consistent with previous studies \citep[][]{Ocvirk2008, Keres2009a, Voort2011a, Correa2018a}. For the accretion onto the halo, we leverage a Bayesian hierarchical formalism to model the continuous evolution of \fcold\ as a function of virial mass and redshift and provide a functional form for the time evolution of the critical mass separating cold and hot-dominated regimes displayed by our simulations.

The outline of this paper is as follows: in \S\ref{sec:simulations}, we summarize the important aspects of our simulations; \S\ref{sec:cold_accretion} is where we define the method employed to compute the cold fraction; we provide a visual inspection of some example galaxies in \S\ref{sec:visual_inspection}; we present and compare the mass accretion rates and cold fractions in \S\ref{sec:mass_accretion_rates} and \S\ref{sec:cold_fraction_final_z}; then we investigate the cosmic evolution of the cold fraction in \S\ref{sec:fcold_cosmic_evolution}; and finally, we provide a concise summary and concluding remarks in \S\ref{sec:conclusion}.

\section{Simulations}\label{sec:simulations}
In this section, we summarise the simulations used for this work, which consist of two different suites split into three total samples. The first suite is the \emph{Numerical Investigation of a Hundred Astrophysical Objects} \citep[NIHAO;][]{Wang2015}, a set of $\sim\!150$ high-resolution cosmological zoom-in simulations covering three orders of magnitude in halo mass $(\sim\!10^9-10^{12}\,\Msun)$ at $z=0$. The second suite is the \emph{High-$z$ Evolution of Large and Luminous Objects} (HELLO) recently introduced in \citet{Waterval2024}, which builds upon NIHAO and consists of currently $\sim\!30$ simulations of massive $(\sim\!10^{12}-10^{13}\,\Msun)$ star-forming galaxies at high-$z$, namely $z=3.6$ and $z=2.0$. In the rest of this manuscript, we differentiate between the two by referring to the former as `HELLOz3.6' and the latter as `HELLOz2.0', respectively. HELLO simulations successfully recover various scaling relations at high-$z$, such as the star-formation main sequence (SFMS) of star-forming galaxies, the stellar-to-halo mass relation, and the size-mass relation.
Both HELLO and NIHAO share mostly the same sub-grid physics which are briefly described below and we refer the reader to the aforementioned references for more details. Differences between the two suites will be made explicit when applicable.

The simulations are based on a flat $\Lambda$CDM model with cosmological parameters as given by \citet{Planck2014}. The Hubble constant $H_0$ is set to 67.1\,$\kms\,\Mpc^{-1}$ and the matter, dark energy, radiation, and baryon densities are, respectively, \{$\Omega_\mathrm{m}$, $\Omega_\Lambda$, $\Omega_\mathrm{r}$, $\Omega_\mathrm{b}$\} = \{0.3175, 0.6824, 0.00008, 0.0490\}. The normalisation of the power spectrum and its slope are $\sigma_8$ = 0.8344 and $n$ = 0.9624.

\subsection{Initial conditions and sub-grid physics}\label{subsec:ic_and_subgrid}

Initial conditions are drawn from a low-resolution DM-only volume simulation, which differs between NIHAO and HELLO. The former use boxes of various sizes (15, 20, and 60 \hMpc) with $400^3$ particles while the latter use haloes at $z=3.6$ and $z=2.0$ from a single volume of size 100 \hMpc\ and $400^3$ particles as well. Haloes are identified with the Amiga Halo Finder \citep[\ahf;][]{Gill2004, Knollmann2009}{}{} and evolved hydrodynamically using the \textsc{\small GASOLINE2} code \citep[][]{Wadsley2017}. We define the virial radius \Rvir\ as the radius containing an average density 200 times times the critical density $\rho_\mathrm{c}(z)$ at the respective redshift. The virial mass \Mvir\ is then the total mass enclosed in a sphere of radius \Rvir\ and is computed as
\begin{equation}
    \Mvir = \frac{4}{3}\,\pi\,\Rvir^3\,200\,\rho_{\mathrm{c}}.
\end{equation}
In the remainder of the paper, we denote all virial quantities with the subscript `vir' and we define the galaxy radius as $\Rgal \equiv 0.2\,\Rvir$.
Each halo of both suites contains about $N_\mathrm{part} \sim 10^6$ within \Rvir\ at their respective redshift.

NIHAO simulations adopt a DM force softening length $\epsilon_\mathrm{DM}$ of 1/40 the inter-particle distance in the highest resolution region from which the gas softening is defined as $\epsilon_\mathrm{gas} = \epsilon_\mathrm{DM}\,\sqrt{\Omega_\mathrm{b}/\Omega_\mathrm{DM}}$, where $\Omega_\mathrm{DM} = \Omega_\mathrm{m} - \Omega_\mathrm{b}$.

Despite a smaller softening, HELLO uses the same threshold for star formation as NIHAO, i.e. $n_{\mathrm{th}} = 10\,\mathrm{cm}^{-3}$, and the temperature threshold is set to $T_{\mathrm{th}} = 15,000$ K. Gas particles with density above $n_{\mathrm{th}}$ and temperature below $T_{\mathrm{th}}$ are allowed to form stars according to the Kennicutt-Schmidt law \citep[][]{Schmidt1959,Kennicutt1998}. Details about the implementation can be found in \citet{Stinson2006,Stinson2013}.

Stellar feedback includes the energy stemming from young bright stars (early stellar feedback; ESF) released as thermal energy as a fraction $\epsilon_{\mathrm{ESF}}=0.13$ of the stellar luminosity, energy from supernovae (SNe) for stars with initial mass $8 < \Mstar < 40\,\Msun$ modelled as the blast wave formalism \citep[][]{Stinson2006} with subsequent radiative cooling turned off ($\sim\!30\,\mathrm{Myr}$) for the affected gas particles, and stellar winds ejecting mass and metals for light stars with $\Mstar < 8\,\Msun$. Both suites adopt a \citet{Chabrier2003} stellar initial mass function.

Chemical enrichment from stars has been updated compared to NIHAO, as per \citet{Buck2021}. On top of SN\,Ia and SN\,II, HELLO includes yields from the asymptotic giant branch (AGB) stars. Yields from these three channels are drawn from \citet{Seitenzahl2013}, \citet{Chieffi2004}, and \citet{Karakas2016}, respectively. All our simulations track the 10 most abundant elements by default (H, He, O, C, Ne, Fe, N, Si, Mg, and S) while HELLO track an additional six elements (Na, Al, Ca, Ti, Sc, and V).

Gas cools via hydrogen, helium, and metal lines and a full description can be found in \citet{Shen2010} and heating from the ultra-violet background (UVB) is modelled as \citet{Haardt2012} in NIHAO. HELLO simulations, on the other hand, include LPF implemented on top of a \citet{FaucherGiguere:2009} UVB, as described in \citet{Obreja2019}.

Finally, black hole (BH) growth and AGN feedback are implemented similarly in all simulations, following the description in \citet{Blank2019}. Haloes reaching a threshold mass of $5 \times 10^{10}\,\Msun$ are seeded with a BH with initial mass $M_{\mathrm{seed}}=10^5\,\Msun$ by converting the gas particle with the lowest gravitational potential. Accretion follows the Bondi model \citep[][]{Bondi1952} with boost parameter $\alpha=70$ and is capped by the Eddington limit \citep[][]{Eddington1921}. During each timestep, the calculated accretion rate determines the mass to be transferred from the most gravitationally-bound gas particle to the BH. The BH luminosity is inferred from the accretion rate and a fraction $\epsilon_{\mathrm{f}}=0.05$ is released as thermal energy to the 50 nearest neighbouring gas particles.

\subsection{Final sample}\label{subsec:final_sample}
Our goal is to compare HELLOz3.6, HELLOz2.0 and NIHAO using galaxies with the closest virial mass at their respective final redshift as possible. For this reason, we select only a subsample of HELLOz3.6 and NIHAO galaxies. The final three samples mentioned at the beginning of \S\ref{sec:simulations} and used in this work, as well as the respective number of galaxies are HELLOz3.6 (10 galaxies), HELLOz2.0 (17 galaxies), and NIHAO (13 galaxies). We display in Table~\ref{tab:general_quantities} the final sample of galaxies with various quantities at the final redshift. The subscript `100' for the star-formation rate (SFR) indicates that it has been averaged over 100 Myr \citep[see][ for details]{Waterval2024}.

\begin{table*}
\centering
\begin{tabular}{ccccccc}
     \toprule
          Simulation &   $z$ & $\Rvir$ & $\log(M_{\mathrm{gas}}(<\Rvir))$ & $\log(\Mstar(<\Rgal))$ & $\log(\Mvir)$ & $\mathrm{SFR}_{100}(<\Rgal)$\\
     \midrule
     g2.47e12 & 3.6 &   87 &  11.4 &  10.5 &  12.3 &  104  \\
     g2.40e12 & 3.6 &   88 &  11.5 &   9.8 &  12.4 &   21  \\
     g2.69e12 & 3.6 &   91 &  11.4 &  10.7 &  12.4 &  158  \\
     g3.76e12 & 3.6 &  102 &  11.7 &  10.5 &  12.6 &  127  \\
     g4.58e12 & 3.6 &   92 &  11.5 &  10.8 &  12.4 &  149  \\
     g2.71e12 & 3.6 &   84 &  11.4 &  10.6 &  12.3 &  123  \\
     g2.49e12 & 3.6 &   91 &  11.5 &  10.5 &  12.4 &  108  \\
     g2.51e12 & 3.6 &   91 &  11.5 &  10.4 &  12.4 &   59  \\
     g2.32e12 & 3.6 &   83 &  11.3 &  10.7 &  12.3 &  125  \\
     g2.96e12 & 3.6 &   95 &  11.6 &  10.5 &  12.5 &  122  \\
     \midrule
     g3.08e12 & 2.0 &  142 &  11.4 &  11.0 & 12.5 &  60 \\
     g3.09e12 & 2.0 &  144 &  11.4 &  11.1 & 12.5 &  65 \\
     g3.00e12 & 2.0 &  135 &  11.4 &  10.8 & 12.4 &  87 \\
     g3.20e12 & 2.0 &  140 &  11.5 &  10.8 & 12.4 &  75 \\
     g2.75e12 & 2.0 &  135 &  11.4 &  11.0 & 12.4 &  66 \\
     g3.03e12 & 2.0 &  119 &  11.2 &  10.9 & 12.2 &  66 \\
     g3.01e12 & 2.0 &  139 &  11.4 &  10.9 & 12.4 &  83 \\
     g2.29e12 & 2.0 &  122 &  11.2 &  10.9 & 12.3 &  81 \\
     g3.35e12 & 2.0 &  120 &  11.3 &  10.4 & 12.2 &  47 \\
     g3.31e12 & 2.0 &  133 &  11.4 &  10.7 & 12.4 &  52 \\
     g3.25e12 & 2.0 &  139 &  11.5 &  10.6 & 12.4 &  46 \\
     g3.38e12 & 2.0 &  146 &  11.5 &  10.9 & 12.5 &  85 \\
     g3.36e12 & 2.0 &  128 &  11.4 &  10.5 & 12.3 &  35 \\
     g2.83e12 & 2.0 &  133 &  11.4 &  10.8 & 12.4 &  63 \\
     g2.63e12 & 2.0 &  133 &  11.4 &  11.0 & 12.4 &  93 \\
     g3.04e12 & 2.0 &  145 &  11.4 &  11.0 & 12.5 & 102 \\
     g2.91e12 & 2.0 &  129 &  11.3 &  10.9 & 12.3 &  40 \\
     \midrule
     g4.84e12 & 0.0 &  336 &  11.3 &  10.9 & 12.6 & 1 \\
     g3.42e12 & 0.0 &  268 &  10.9 &  10.9 & 12.3 & 0 \\
     g1.62e12 & 0.0 &  212 &  10.8 &  10.3 & 12.0 & 1 \\
     g1.27e12 & 0.0 &  203 &  10.4 &  10.6 & 11.9 & 0 \\
     g1.26e12 & 0.0 &  217 &  10.8 &  10.5 & 12.0 & 2 \\
     g4.81e12 & 0.0 &  351 &  11.3 &  11.0 & 12.7 & 0 \\
     g4.41e12 & 0.0 &  325 &  10.4 &  10.8 & 12.6 & 0 \\
     g2.37e12 & 0.0 &  268 &  10.8 &  10.7 & 12.3 & 0 \\
     g2.71e12 & 0.0 &  277 &  11.1 &  10.6 & 12.3 & 1 \\
     g3.74e12 & 0.0 &  312 &  11.2 &  10.8 & 12.5 & 1 \\
     g5.22e12 & 0.0 &  354 &  11.4 &  11.0 & 12.7 & 0 \\
     g1.55e12 & 0.0 &  222 &  10.8 &  10.6 & 12.1 & 0 \\
     g4.55e12 & 0.0 &  337 &  11.3 &  10.9 & 12.6 & 0 \\
     \bottomrule
     \end{tabular}
     \label{tab:general_quantities}
     \caption{Galaxy sample used in this work. The columns from left to right are redshift, virial radius, gas mass inside the halo, galaxy stellar mass, halo virial mass, and star-formation rate. All distances and masses are in units of kpc and \Msun, respectively, while SFRs are in units of $\Msun\,\mathrm{yr}^{-1}$.}
\end{table*}

\section{Cold accretion}\label{sec:cold_accretion}
Various authors in the literature use different definitions for i) what is considered accretion, ii) how it is calculated, and iii) the mode (hot or cold) in which the accretion takes place. Hereafter we briefly summarise these differences.

First, accretion is generally differentiated whether it is total, i.e. \textit{all} accreted gas, or \textit{smooth}, i.e. disregarding mergers. Simulations of galaxy formation indicate that galaxies predominantly grow via smooth accretion \citep[e.g.][]{Murali2002}, while merger-driven growth only becomes relevant for groups and clusters \citep[][]{Voort2011a}. The distinction between smooth and total accretion is fundamental when, e.g., comparing accretion rates since total accretion rates are naturally larger.

Second, in simulations, accretion and accretion rates can be either defined in an \textit{Eulerian} way (instantaneous) or in a \textit{Lagrangian} way (over two successive timesteps). In the former, accretion through some boundary (e.g. \Rvir) is calculated within a shell of an arbitrary thickness around the boundary by considering the (smooth) gas present within the shell and using its radial velocity to determine the mass flow through it (see, \S\ref{subsec:lagrangian_mass_flows}). In the latter, accretion is calculated by tracking gas particles through consecutive snapshots and the rate can be obtained by averaging over the timestep between the two snapshots.

Third, there are mainly three different approaches to define the mode of the accreted gas onto haloes and galaxies: a fixed temperature cutoff \Tcutoff, a temperature cutoff expressed as a fraction of the virial temperature \Tvir, or an entropy criterion. As first demonstrated by \citet{Keres2005}, $\Tmax / \Tvir$ does not appear to be a suitable method, and they thus choose a fixed temperature threshold of $2.5 \times 10^{5}$ K, which has been widely used by other authors \citep[e.g.][]{Ocvirk2008,Keres2009a,FaucherGiguere2011,Voort2011a, Correa2018a}. In fact, after the gas encounters a virial shock, the post-shock temperature is not necessarily expected to reach \Tvir\ but $\gtrsim (3/8)\,\Tvir$ \citep[][]{Dekel2006}, rendering $\Tcutoff = \Tvir$ too conservative.
Note that previous authors have investigated the dependence of the cold fraction \fcold\ on the specific definition used to discriminate between the two modes \citep[e.g.][]{Voort2011a,Nelson2013,Correa2018a} and found that the inferred cold fractions greatly depend upon the chosen criterion, albeit mostly for haloes below $10^{12}\,\Msun$.
Finally, \citet{Brooks2009} use the entropy of the smoothly accreted gas to classify whether gas particles have undergone a shock. They compare the entropy criterion with $\Tcutoff = 2.5 \times 10^{5}$ K and find that the shocked fraction obtained from the former differs from the hot fraction resulting from the latter by less than 7~per~cent and thus conclude that the two definitions are equivalent.

In this work, we adopt a fiducial definition of \textit{smooth} accretion using the \textit{Lagrangian} method and a temperature cutoff $\Tcutoff = 2.5 \times 10^{5}$ K, applied over the entire history of the relevant gas particles, which we describe in more details below.

\subsection{Tracing gas particles}\label{subsec:gas_tracks}
To classify gas particles as being accreted in the cold or hot mode, we trace the history of the state of each particle deemed as accreted across all previous snapshots. We study the mode of accretion at two relevant radii: \Rvir\ and \Rgal, defining accretion onto the halo and galaxy, respectively.

At the target redshift of choice, we select all smooth particles within $0 < r < \Rgal$ and $\Rgal < r < \Rvir$ that were outside these regions in the snapshot before, where the timestep between two contiguous snapshots is $\sim 200$ Myr. We define a particle as smooth if it has never been bound to any halo, with the exception of the main one. In other words, we select all smooth gas particles that entered the galaxy (halo) during the last timestep and trace them back across all previous snapshots, until the initial one. This process is then repeated for each galaxy in all three samples (HELLOz3.6, HELLOz2.0, and NIHAO) and the stored historical values of their gas temperature can then be used to obtain the cold fraction accreted onto the galaxy and the halo. Each particle for which $\Tmax < \Tcutoff$ is classified as cold and \fcold\ is then calculated as
\begin{equation}
    \fcold = \frac{\sum m_{i,\mathrm{cold}}}{\sum m_j},
\end{equation}
where $m$ is the mass of the gas particle at the final timestep and $i$ ($j$) runs over cold (all) accreted particles.

\begin{figure*}
\centering
\includegraphics[width=\linewidth]{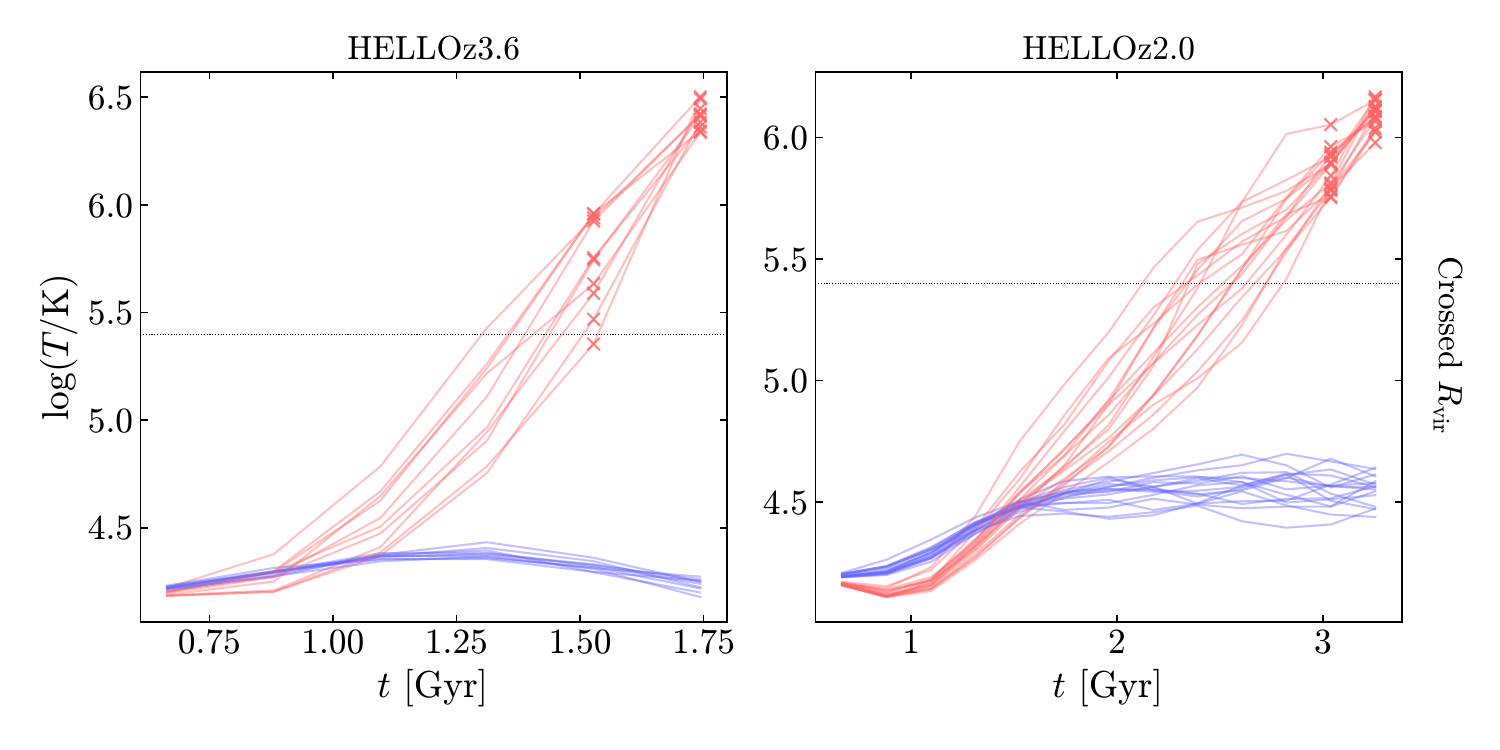}
\caption{Mass-weighted temperature average of the particles accreted onto \Rvir\ in the hot mode (red) and cold mode (blue) calculated for each galaxy at each snapshot. The left (right) panel shows the results for HELLOz3.6 (HELLOz2.0) galaxies. The horizontal dashed line indicates \Tcutoff\ and the x's in the hot tracks mark the two last snapshots.}
\label{fig:temp_tracks_vs_time}
\end{figure*}

Using the history of the entire particle is rather conservative and some authors adopt a lookback time threshold beyond which the state of the particle is not considered \citep[e.g.][]{Brooks2009} but as Fig.~\ref{fig:temp_tracks_vs_time} demonstrates,
the average temperature of particles in the hot mode in high-$z$ increases very smoothly over time. Fig.~\ref{fig:temp_tracks_vs_time} shows for each galaxy in HELLOz3.6 (left panel) and HELLOz2.0 (right panel) the mass-weighted average of the temperature tracks of particles classified in the hot mode (red) and cold mode (blue) that crossed \Rvir\ between the last two snapshots, which are indicated by `x' markers in the hot tracks. The horizontal dashed line represents \Tcutoff. While the hot tracks can cross \Tcutoff\ a few hundred Myr before entering the halo, a significant amount of gas that would later cool and be accreted cold would likely manifest itself in Fig~\ref{fig:temp_tracks_vs_time}. Moreover, particles crossing \Tcutoff\ within the IGM find themselves in a very low-density environment, thus greatly limiting their ability to cool efficiently until they reach the high-density environment of the halo. Contamination by a large amount of particles that crossed \Tcutoff\ and cooled back before entering the halo is thereby highly unlikely. The situation is slightly more convoluted for NIHAO galaxies at $z=0$, since a substantial amount of gas particles potentially crossed \Tcutoff\ at early times, owing to their significantly longer evolution time and possibly implying multiple cycles of accretion and outflows from the same particle \citep[][]{Oppenheimer2010, Woods2014, Ubler2014, Tollet2019, Tollet2022}. However, as shown in Appendix~\ref{app:cold_frac_evol}, the different methods of calculating the cold fraction overall impact the accretion onto the galaxy more dramatically than the accretion onto the halo, which is ultimately the focus of our attempt to model the \fcold evolution with mass and redshift in Section~\ref{sec:fcold_cosmic_evolution}.

\subsection{Instantaneous accretion rates}\label{subsec:lagrangian_mass_flows}
We conclude this section by defining how we obtain instantaneous mass accretion rates in this work.
Mathematically, the accretion rate of an arbitrary element through an infinitesimally thin spherical shell $S$ can be calculated by integrating the element's density $\rho$ and velocity $\mathbf{v}$ over the shell
\begin{equation}
    \dot{M} = \int_S \rho\,\mathbfit{v}\cdot\mathrm{d}\mathbfit{S} = \int_S \rho\,v_{r}\,\mathrm{d}S,
\end{equation}
where $v_r$ denotes the radial velocity.
Given that our simulations contain a set of discrete particles $i$ with masses $m_i$, the integral can be approximated by a summation evaluated over a shell of thickness $\Delta r$

\begin{equation}\label{eq:accretion_rate}
    \dot{M} = \sum_j m_j\,v_{r,j}\,\frac{1}{\Delta r}.
\end{equation}
In Eq.~\ref{eq:accretion_rate} above, the summation is performed over all particles $j$ within the shell.

\section{Visual inspection}\label{sec:visual_inspection}
In this section, we begin with a qualitative visual inspection of three example galaxies, one from each sample: g3.76e12 (HELLOz3.6), g3.31e12 (HELLOz2.0), and g3.42e12 (NIHAO). All three galaxies have roughly the same halo mass $\sim\!3 \times 10^{12}\,\Msun$. In \S\ref{subsec:maps}, we present maps of the distribution of DM, gas, gas temperature, and gas radial velocity, while in \S\ref{subsec:gas_phase_space}, we display the gas temperature-density diagrams, binned in $\log(n)$ and $\log(T)$ and colour-coded in mass and radial velocity. Note that in this section we do not distinguish between smooth and bound particles and include all of them.

\subsection{Large-scale maps}\label{subsec:maps}
\begin{figure*}
\centering
\includegraphics[width=\linewidth]{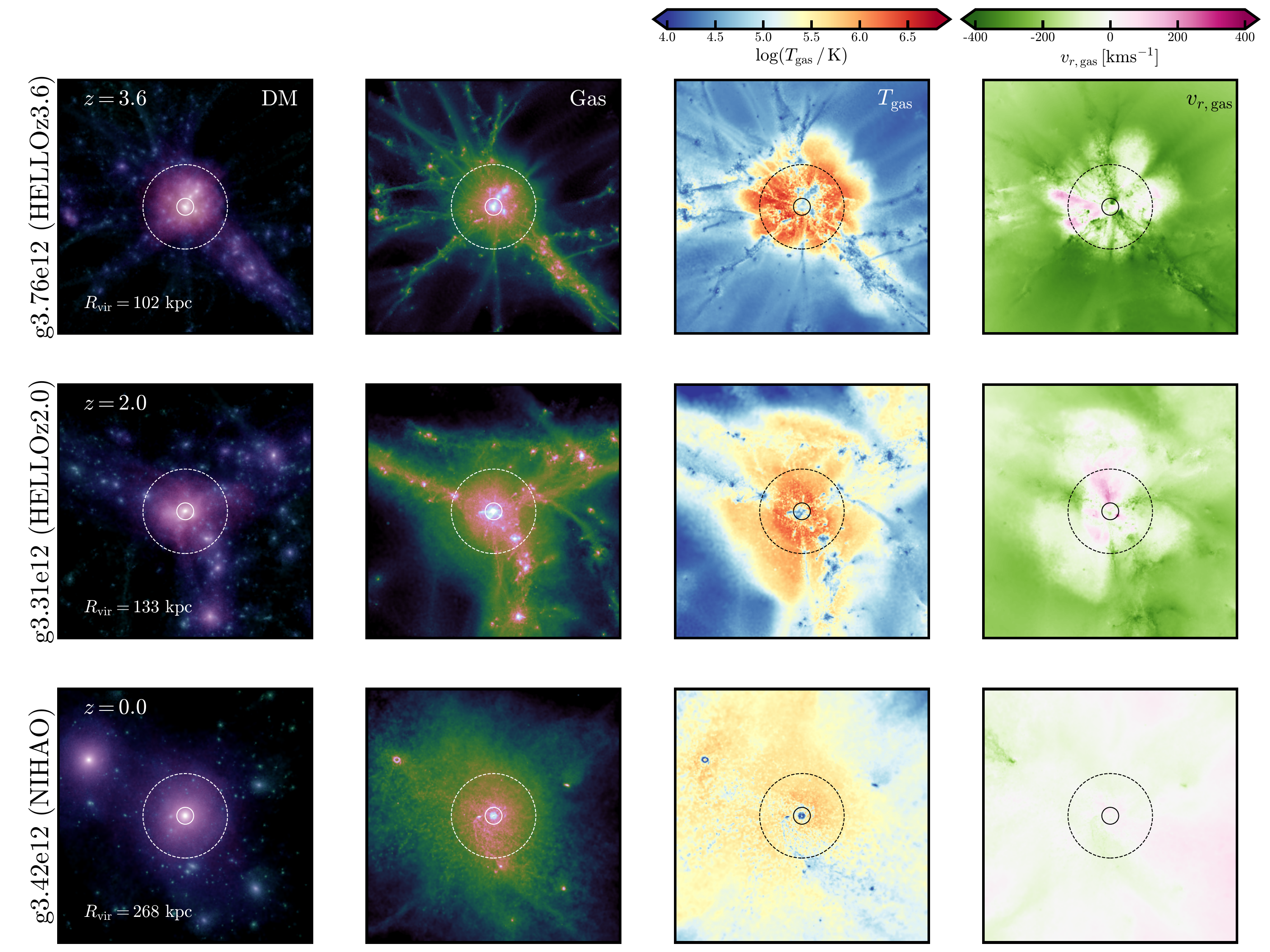}
\caption{Large-scale maps of the distribution of DM, gas, gas temperature and gas radial velocity for three example galaxies, one from each sample (HELLOz3.6, HELLOz2.0, and NIHAO, from top to bottom). The first column is an RGB image of the DM where the hue is associated with the velocity dispersion and the brightness to the mass surface density. The second column shows the gas surface density. The third and fourth columns are the mass-weighted average log temperature and radial velocity, respectively. In each panel, the small and large circles delimit areas of radii \Rgal\ and \Rvir, and each map is made of $500 \times 500$ pixels, covering a $6\,\Rvir \times 6\,\Rvir$ region centered on the galaxy. All maps are made from \textit{all} particles, i.e. smooth and bound.}
\label{fig:large_scale_maps}
\end{figure*}
Fig.~\ref{fig:large_scale_maps} shows the distributions of DM, gas, gas temperature, and gas radial velocity from left to right for the three aforementioned example galaxies (top to bottom). The DM maps are RGB images whose brightness corresponds to the mass surface density and the hue maps to the velocity dispersion. The second column shows the gas mass surface density. The last two columns exhibit the average mass-weighted temperature and radial velocity, respectively.
In each panel, the small and large circles delimit areas of radii \Rgal\ and \Rvir, respectively, and each map is made of $500 \times 500$ pixels, covering a $6\,\Rvir \times 6\,\Rvir$ region centered on the galaxy. These maps show \textit{all} particles, i.e. both smooth and bound ones.
Note that in both temperature and velocity maps, the colour range is kept constant across the three galaxies. This highlights the strong bimodality in cold/hot and inflows/outflows in HELLO galaxies, while the NIHAO example exhibits a more homogeneous distribution around $\Tcutoff$ and $v_r = 0$. For the DM and gas maps, on the other hand, the colour gradient is adapted each time, in order to highlight the halo internal and neighbouring structures rather than to compare the densities themselves.
Before describing the maps, we emphasise that their purpose is for visual inspection only and are thus qualitative in nature.

The most striking difference between the three examples presented is the filamentary structures clearly visible at high-$z$ and completely absent at $z=0$. These three galaxies have very similar halo masses ($2{-}4\times 10^{12}\,\Msun$) but at three different redshifts spanning over 10 Gyr. As previously mentioned, haloes around $\sim\!10^{12}$ at high-$z$ are rare high-density peaks who tends to be nodes intersecting multiple filaments \citep[][]{Bond1996,Springel2005a}. On the other hand at late times, say $z \lesssim 1$, similar mass haloes are fairly typical and embedded in larger structures. From Fig.~1 in \citet{Mandelker2018} for example, the redshift mass evolution of $2\sigma$ density fluctuations shows that for a $10^{12}\,\Msun$ halo, the corresponding redshift is $z \sim 3$. Recent results from \citet{GalarragaEspinosa2024} using the MilleniumTNG simulation suite \citep[][]{HernandezAguayo2023, Pakmor2023} indicate that the mean of the mass function of nodes connected to filaments is $\log(\Mvir/\Msun) = 12.06$ at $z=4$, 12.47 at $z=3$, 12.9 at $z=2$ and 13.66 at $z=0$ (their Fig.~5). Moreover, they also study the evolution of the connectivity (i.e. the number of connected filaments) and find that it increases with redshift. From these results, our HELLOz3.6 galaxies are at the centre of nodes, while HELLOz2.0 haloes are slightly below the expected node mass at $z=2$.
NIHAO galaxies, however, find themselves in haloes of mass an order of magnitude below the one of typical present-day nodes. Qualitatively, our plot suggests that for these example galaxies, there are less filaments attached to the halo at $z=2$ compared to $z=3.6$ and that they appear larger, consistent with the growth of large-scale structures in an expanding Universe. The NIHAO DM map (bottom left) shows some continuous large-scale structures extending from the top left to the bottom right, which could be a large filament embedding the halo.

Moving on to the gas temperature and radial velocity (two left columns), the difference between high- and low-$z$ is once again striking. Starting with $z=3.6$, a clear bimodality is visible in the gas temperature, with the cold gas filling the diffuse IGM as well as higher-density filaments streaming towards the central galaxy. Extending slightly beyond \Rvir, plumes of hot outflowing gas fill the low-density environment between filaments. These outflows are the direct consequence of the high SFR currently found in the galaxy (127\,\Msun, see Fig.~\ref{fig:hello_nihao_temp_dens}). The galaxy at $z=2$ broadly exhibits similar characteristics, albeit with a more extended hot component. Hot outflows again fill the regions around cold filaments inside the halo, while beyond \Rvir\ both phases seem to start mixing as far as $\sim2\Rvir$, indicated by the relatively lower temperature and low radial velocities. Finally, at $z=0$, both temperature and radial velocity maps exhibit a rather homogeneous and static distribution, with cold inflows tied to nearby incoming satellites.

\subsection{Gas phase space}\label{subsec:gas_phase_space}
\begin{figure*}
\centering
\includegraphics[width=\linewidth]{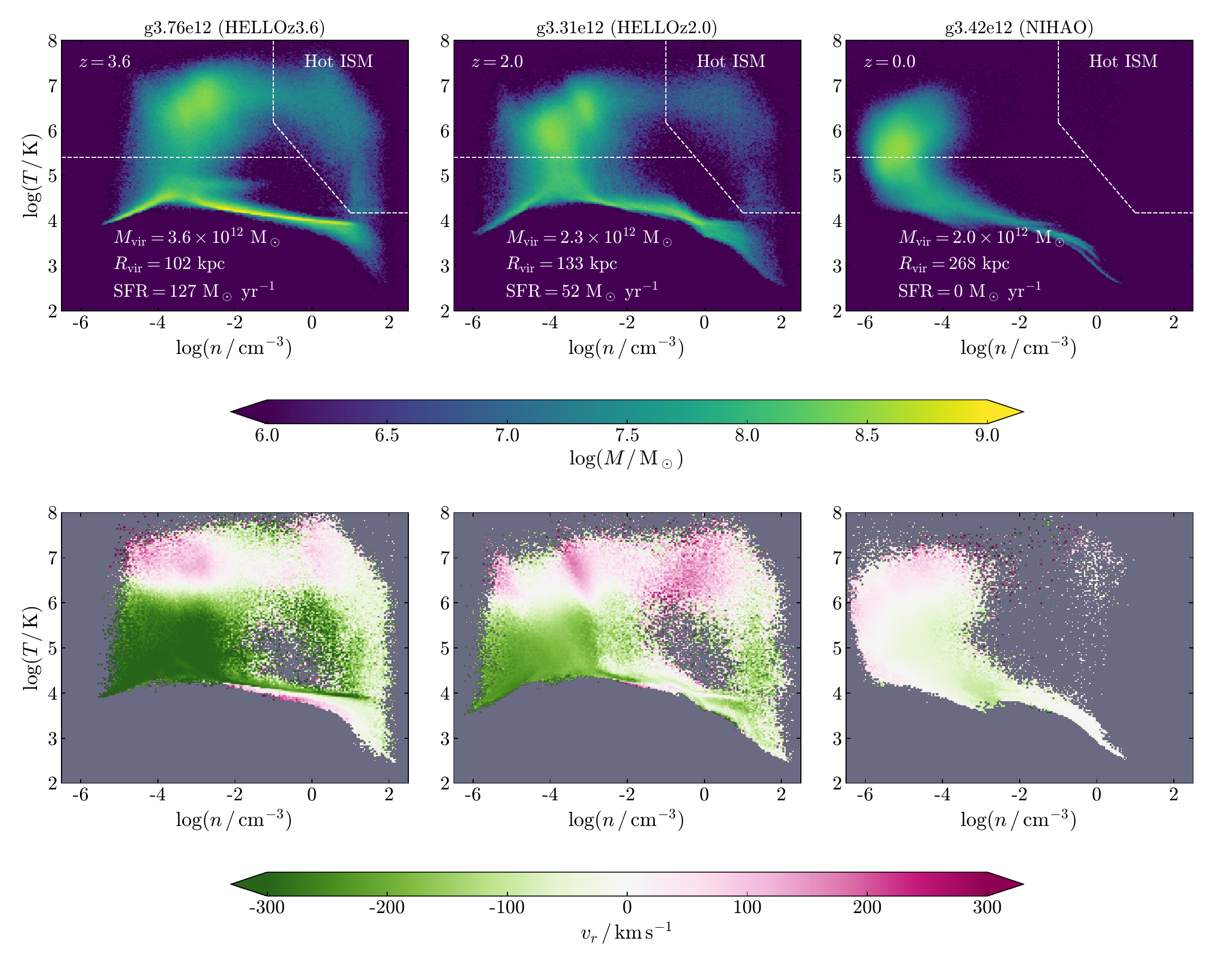}
\caption{Temperature-density diagrams of three example galaxies (same as Fig.~\ref{fig:large_scale_maps}, one from each sample, i.e. HELLOz3.6, HELLOz2.0, and NIHAO, from left to right). Each panel is a 200x200 pixel map binned in $\log(T)$ and $\log(n)$ and contains gas particles within $0 < r < 3\,\Rvir$. The colour map displays the total mass per bin in the upper row and the mass weighted average radial velocity in the lower row. The horizontal dashed line in the upper panels marks the $\Tcutoff = 2.5 \times 10^{5}$ K above and below which we identify the hot- and cold-gas phases, while the polygonal in the upper-right corner separates the hot ISM region from the CGM/IGM-dominated region. All maps are made from \textit{all} particles, i.e. smooth and bound.}
\label{fig:hello_nihao_temp_dens}
\end{figure*}
Here, we begin by identifying the multiple components of the gas phase space in and around the host halo of the same three example galaxies as in \S\ref{subsec:maps}. Fig.~\ref{fig:hello_nihao_temp_dens} shows the temperature-density diagrams for these galaxies with similar virial mass, one from each sample, at $z=3.6$, $z=2$ and $z=0$ from left to right. These diagrams contain the distribution of gas particles within a sphere extending to $3\,\Rvir$ around the center of the galaxy. The respective redshift is shown alongside the halo mass,the virial radius, and SFR in each case. The upper panels show the total mass of the gas in each bin, while the lower panels show the mass-weighted radial velocity average of the gas particles in each bin.

In each row, we use the same colour bar for all three galaxies in order to highlight the differences between the simulations. Temperatures and densities are binned using 200 bins per side. Visual cues are added to help distinguish between the different gas phases and accretion modes. Particles in the hot interstellar medium (ISM) are identified by having a temperature greater than the threshold for star formation (15,000 K), a pressure above the typical pressure of the ISM, $p_{\mathrm{ISM}}$, displayed as a diagonal isobar, and a density $n > -1$ cm$^{-3}$. Note that this definition of the hot ISM is merely qualitative but sufficient for the purpose of the figure, which is to provide an approximate overview of the differences in gas phases between our three selected galaxies. Finally, the line at $2.5 \times 10^5$ K represents \Tcutoff\ to separate the cold and hot gas. 

Starting by looking at all three upper panels together, the main striking difference is the amount of gas in the hot ISM. NIHAO galaxy at $z=0$ on the right is completely devoid of gas in this phase, owing to its vanishing SFR of $0\,\Msun\,\mathrm{yr}^{-1}$. Both HELLO galaxies, on the other hand, have their hot ISM filled with a significant amount of gas, which is a direct consequence of their large SFRs reaching 127 and $52\,\Msun\,\mathrm{yr}^{-1}$ at $z=3.6$ and $z=2$, respectively, and the ensuing feedback energy released. 

In the bottom row, we retrieve the general characteristics of inflows and outflows previously observed in Fig.~\ref{fig:large_scale_maps}. Interestingly, both HELLO galaxies exhibit some hot accretion of gas with temperature $T \sim 10^6$ K. These are likely made from outflows being re-accreted and/or incoming diffuse gas heated when encountering the hot outflows. In light of that, we investigated the hot incoming gas within the ISM of g3.76e12 (lower left panel in Fig.~\ref{fig:hello_nihao_temp_dens}, $0 < \log(n / \mathrm{cm}^{-3}) < 1$ and $5 < \log(T/\mathrm{K}) < 6$): we found that, in the two preceding snapshots, all these particles exhibited positive radial velocities, indicating they were outflowing. This suggests that the gas is being re-accreted after being heated by supernova feedback. Finally, similarly to the radial velocity maps in Fig.~\ref{fig:large_scale_maps}, the gas around the NIHAO halo shows very low accretion and outflows compared to both HELLO examples.

\section{Instantaneous mass accretion rates}\label{sec:mass_accretion_rates}
\begin{figure*}
\centering
\includegraphics[width=\linewidth]{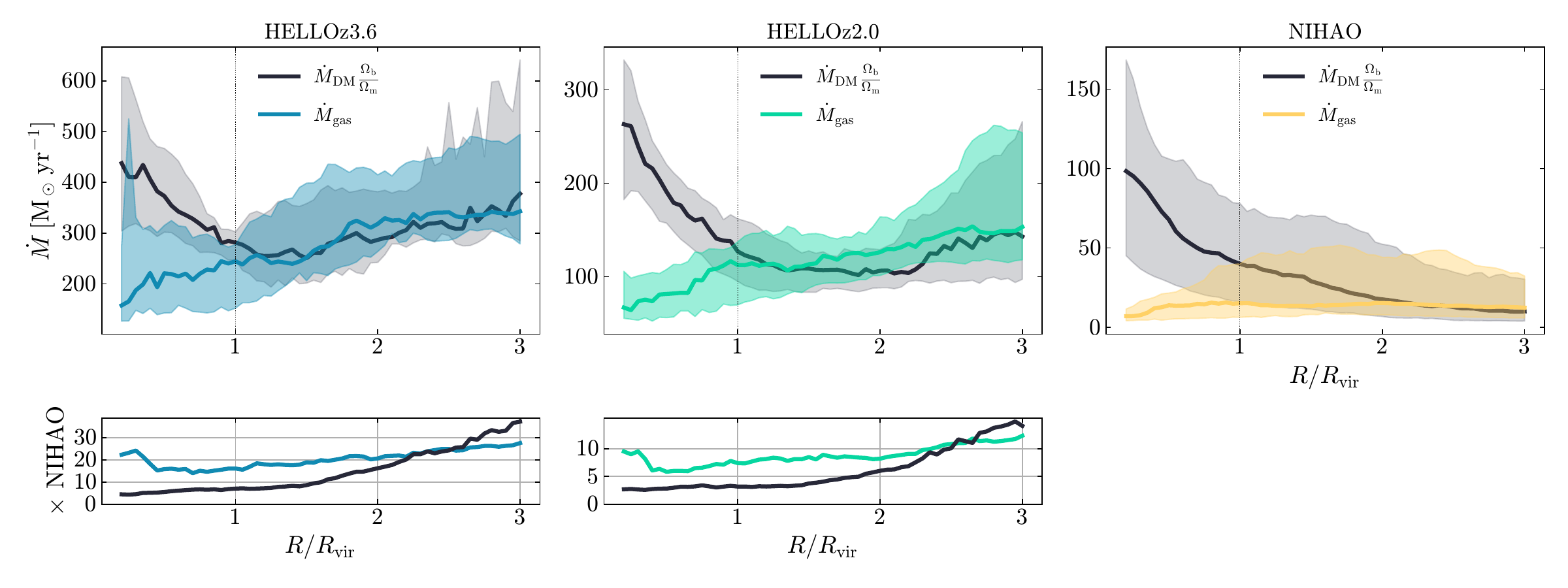}
\caption{Smooth inflowing radial mass accretion rates of DM (scaled by $\Omegab / \OmegaM$) and gas for HELLOz3.6 (left), HELLOz2.0 (middle), and NIHAO (right) galaxies. The continuous curves represent the median rates from all galaxies within the respective sample, while the shaded regions delimit the 16th and 84th percentiles of each distribution. The rates are calculated in spherical shells of width $0.05\,\Rvir$ and extending from $\Rgal \equiv 0.2\,\Rvir$ to $3\,\Rvir$. The lower panels show how HELLO median accretion rates compare to NIHAO.}
\label{fig:mdot_dm_vs_gas}
\end{figure*}
Fig.~\ref{fig:mdot_dm_vs_gas} exhibits the radial inflowing median mass accretion rates of smooth gas (coloured curves) and DM (black curves) particles for our three samples at $z=3.6$, $z=2.0$, and $z=0$, expressed as a fraction of \Rvir\ from 0.2 to 3. The accretion rates are calculated using shells of thickness 0.05\,\Rvir\ following Eq.~\ref{eq:accretion_rate}. The shaded regions delimit the 16th and 84th percentiles and the vertical dotted line shows the virial radius, $R / \Rvir = 1$. The DM accretion rate is scaled by the baryon fraction $\Omegab / \OmegaM$ to distinguish where the gas accretion decouples from free-fall. The lower panels for the two HELLO cases show the fraction of accretion rate with respect to NIHAO galaxies at $z=0$.

All three samples present overall a similar behaviour. Outside of the halo, the gas accretion is coupled to the DM accretion until a certain point where both tracks begin to diverge. For HELLO galaxies, this happens around $R/\Rvir=1$, below which the DM particles fall into the steep potential of the halo and are accelerated towards the centre of the halo. Gas particles, on the other hand, are subject to shock heating, thermal pressure, ram pressure, and SN or AGN-driven outflows which can all slow them down, thus reducing the mass accretion rate. The decoupling between gas and DM accretion rates happens earlier around $R/\Rvir=2$ in NIHAO galaxies at $z=0$, indicating that a hot halo component has settled beyond the virial radius, due to these galaxies having evolved for $\sim 10$ Gyr longer than their HELLO counterparts.

As for the actual numbers, accretion rates in present-day galaxies are significantly lower than at high-$z$. NIHAO galaxies exhibit almost no accretion ($\lesssim 20\,\mathrm{\Msun\,yr}^{-1}$) out to $3\,\Rvir$ for both the DM and gas components. Moreover, this value remains roughly constant down to the \Rgal\ for the gas. The DM accretion rate increases to $\sim 50\,\mathrm{\Msun\,yr}^{-1}$ at \Rvir\ and doubles to reach $\sim 100\,\mathrm{\Msun\,yr}^{-1}$ at \Rgal. At $z=2$ ($z=3.6$), HELLO galaxies display accretion rates roughly 3 (6) and 7.5 (15) times higher at \Rvir\ for DM and gas, respectively. These correspond to rates of $\sim 100$ and $\sim 300\,\mathrm{\Msun\,yr}^{-1}$ at $z=2$ and $z=3.6$. Inside the halo, the DM accretion rate ratio between HELLO and NIHAO remains practically constant, owing to their similar virial mass. While all three samples show diminishing gas accretion rates from \Rvir\ down to \Rgal, HELLO galaxies exhibit an increase of inflowing gas compared to NIHAO, where the relative fraction grows from 15 to 20 and 5 to 10 for HELLOz3.6 and HELLOz2.0, respectively.

\begin{figure}
\centering
\includegraphics[width=\columnwidth]{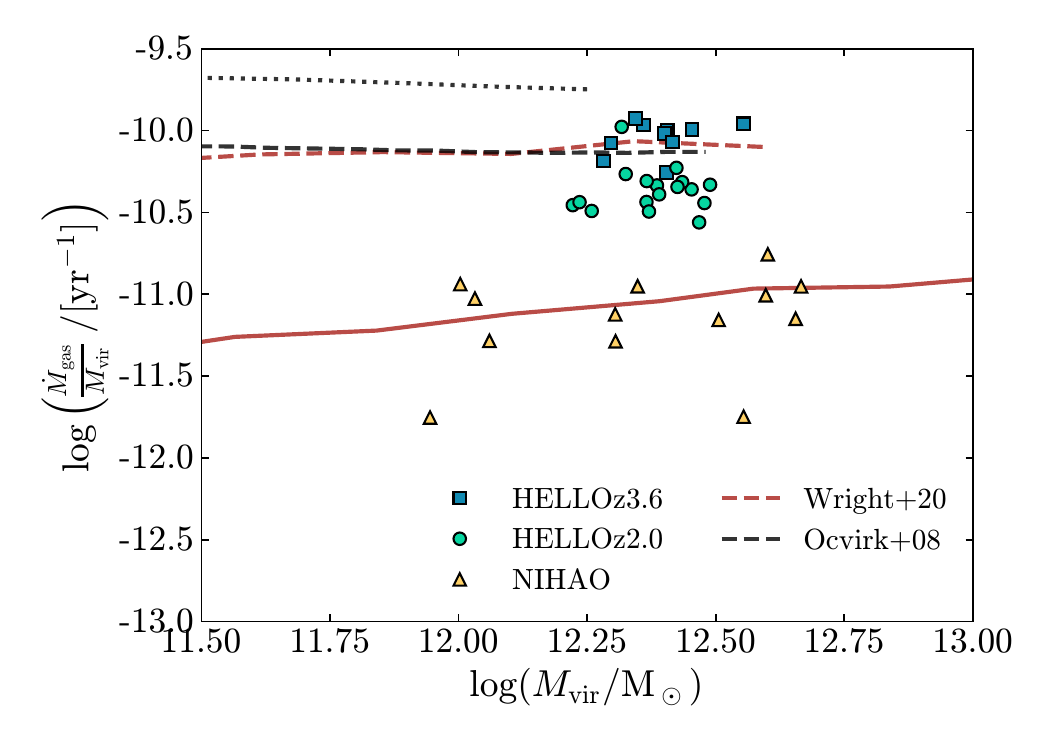}
\caption{Specific inflowing smooth gas accretion rates onto the halo for our galaxies (squares, HELLOz3.6; circles, HELLOz2.0; triangles, NIHAO) calculated within a shell encompassing 0.95--1\,\Rvir. We compare our galaxies to previous results from the literature \citep[][]{Ocvirk2008,Wright2020} at different redshifts, distinguished by their linestyles ($z=0$, continuous; $z=2$, dashed; $z=4$, dotted).}
\label{fig:mdot_comparison}
\end{figure}
We compare the specific gas accretion rates onto the halo of our sample (squares, HELLOz3.6; circles, HELLOz2.0; triangles, NIHAO) with simulations from previous works and using a similar method \citep[][black and red, respectively]{Ocvirk2008, Wright2020} in Fig.~\ref{fig:mdot_comparison}. Different line styles represent different redshifts, namely $z=0$ (continuous), $z=2$ (dashed), and $z=4$ (dotted).
The specific gas accretion rate is simply the gas accretion rate scaled by the inverse virial mass, $\dot{M}_{\mathrm{gas}} / \Mvir$, and we express it in units of $\mathrm{yr}^{-1}$.

NIHAO and HELLO galaxies are generally in qualitative agreement with other simulations within 0.5 dex, albeit at high-$z$ where some slight discrepancy appears. At $z=2$, HELLOz2.0 galaxies deviate from \citet{Ocvirk2008} and \citet{Wright2020} estimates by $\sim\!0.3$ dex. While the relation from \citet{Ocvirk2008} does not extend past $\Mvir \sim 10^{12}\,\Msun$ at $z=4$ (dotted lines), HELLOz3.6 simulations seem to slightly underpredict it as well.

\citet{Ocvirk2008} analyse simulations from the Horizon-MareNostrum suite, which uses the Adaptive Mesh Refinement (AMR) code \textsc{\small RAMSES} \citep[][]{Teyssier2002} and incorporates stellar but no AGN feedback. Furthermore, they use a density-based criterion to separate diffuse accretion from satellites by removing cells exceeding their star-formation threshold, while we remove all particles bound to a DM halo. 
\citet{Wright2020} use galaxies from the EAGLE simulations \citep[][]{Crain2015,Schaye2015}, which incorporates both stellar and AGN feedback. Their preferred method for defining mass accretion rates is Lagrangian, but we show their `Total \Rtwohun\ Inflow' result as it is Eulerian and thus more consistent for comparison. This method includes, however, all particles and is therefore expected to lead to higher values than the ones obtained from our smooth accretion rates. We note that while NIHAO galaxies exhibit a larger scatter than their HELLO counterparts, they are nonetheless in good agreement with \citet{Wright2020} and do not qualitatively show signs of a normalisation bias.
As shown by Fig.~\ref{fig:large_scale_maps} for the example galaxy g3.42e12, its surroundings are relatively homogeneous, hot, and static, likely reducing the importance of the distinction between accretion from mergers and diffuse particles.

\begin{figure*}
\centering
\includegraphics[width=\linewidth]{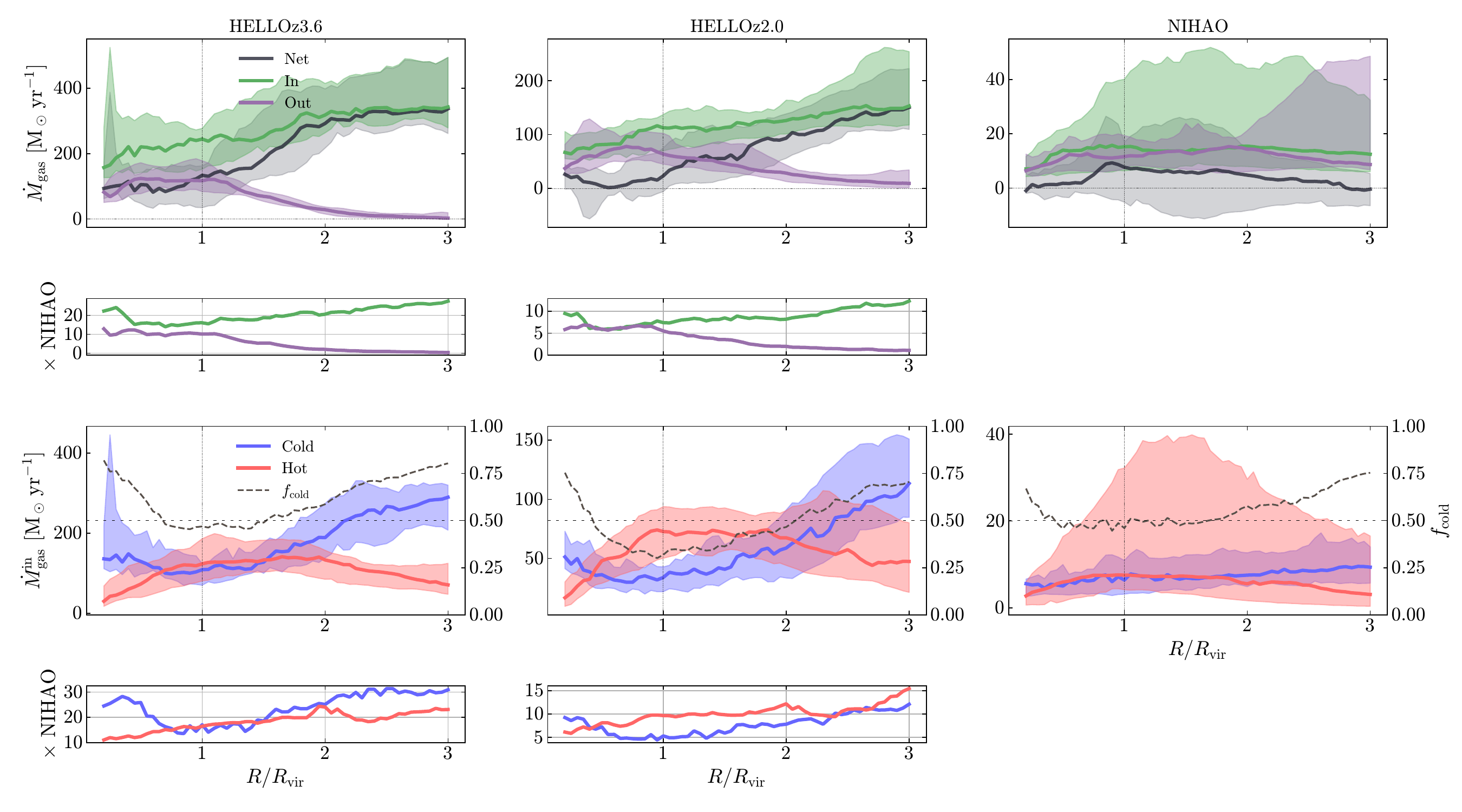}
\caption{Smooth gas mass accretion rates for HELLOz3.6 (left), HELLOz2.0 (middle), and NIHAO (right) separated in three components (top panels): inflowing gas (green), outflowing gas (purple) and net ($\mathrm{in}-\mathrm{out}$; black). Shaded regions encompass the 16th and 84th percentiles. The thin dotted lines indicate the locations of \Rvir (vertical) and $\dot{M}_{\mathrm{gas}}=0$ (horizontal). The lower panels show the inflowing rates, split between cold and hot accretion rates, where the distinction is made with respect to the current gas temperature. Superimposed to the accretion rates is the corresponding \fcold, simply obtained by dividing the cold accretion rate by the total inflowing rate. The horizontal dashed line marks the transition between hot- and cold-dominated gas, i.e. $\fcold=0.5$. All rates are calculated in spherical shells $0.05\,\Rvir$ thick, from \Rgal\ to $3\Rvir$. Similar to Fig.~\ref{fig:mdot_dm_vs_gas}, the intermediate smaller panels show the comparison between HELLO and NIHAO.}
\label{fig:mdot_gas}
\end{figure*}
We end this section by looking at the gas mass accretion rates in more detail in Fig.~\ref{fig:mdot_gas}. The upper panels showcase the median $\dot{M}_{\mathrm{gas}}$ for HELLOz3.6 (left), HELLOz2.0 (middle), and NIHAO (right) separated in three components: inflowing gas (green), outflowing gas (purple) and net ($\mathrm{in}-\mathrm{out}$; black). The green curves correspond to the coloured curve in Fig.~\ref{fig:mdot_dm_vs_gas}. Shaded regions encompass the 16th and 84th percentiles. The thin dotted lines indicate the location of \Rvir (vertical) and where $\dot{M}_{\mathrm{gas}}=0$ (horizontal). The lower panels focus on the inflows only, split between cold (blue) and hot (red), using \Tcutoff\ applied to the current gas temperature as the boundary between the two modes. Here again, the virial radius is indicated by the thin vertical dotted line. Superimposed to the accretion rates is the corresponding \fcold, simply obtained by dividing the cold accretion rate by the total inflowing rate. The horizontal dashed line marks the transition between hot and cold dominated, i.e. $\fcold=0.5$. All rates are calculated in spherical shells $0.05\,\Rvir$ thick, from \Rgal to $3\Rvir$.

Beginning with the upper panels of Fig.~\ref{fig:mdot_gas}, accretion rates between present-day and high-$z$ galaxies span roughly an order of magnitude. NIHAO galaxies exhibit approximately an equal amount of gas flowing in and out. The net rate on average never exceeds $15\,\Msun\,\mathrm{yr}^{-1}$ and inflows/outflows remain below $20\,\Msun\,\mathrm{yr}^{-1}$ at all radii. Accretion rates are much higher at higher $z$. At \Rvir, HELLOz2.0 exhibit inflows of $~110\,\Msun\,\mathrm{yr}^{-1}$ which decreases to $\sim\!70\,\Msun\,\mathrm{yr}^{-1}$ at \Rgal. HELLOz3.6 shows a similar trend, but even with higher values, i.e. $~240\,\Msun\,\mathrm{yr}^{-1}$ at \Rvir\ down to $~160 \Msun\,\mathrm{yr}^{-1}$ at \Rgal. The main difference between HELLOz2.0 and HELLOz3.6 is that the former display comparable outflowing and inflowing rates, leading to a low net rate $\lesssim 20\,\Msun\,\mathrm{yr}^{-1}$ between \Rgal\ and \Rvir, even reaching almost zero around $0.5\,\Rvir$. The latter, however, maintain a net rate of $\sim\!100\,\Msun\,\mathrm{yr}^{-1}$ with an equivalent outflow rate. In summary, while nothing much happens around NIHAO galaxies, HELLO galaxies enjoy significantly higher gas accretion rates (and outflows). These outflows are roughly matching the inflows at $z=2$ but are subdominant at $z=3.6$. As before, the panels below the in-and-out radial distributions show how HELLO galaxies compare to NIHAO.

Moving on to hot and cold accretion (bottom panels of Fig.~\ref{fig:mdot_gas}), all three panels broadly display the same behaviour. Far from the halo ($R/\Rvir \sim 3$), 70--80~per~cent of the incoming gas is below \Tcutoff. This fraction reduces towards $\sim\!$1--2 \Rvir\ where the hot accretion rate peaks and \fcold decreases to 50~per~cent for HELLOz3.6 and NIHAO, and 30~per~cent for HELLOz2.0. The likely explanation is that at $z=3.6$, while the galaxies are (more) actively forming stars and thus releasing more feedback energy, the presence of multiple cold streams embedded in filaments (see Fig.~\ref{fig:large_scale_maps}), can maintain a relatively high inflow of cold gas. At $z=2$ on the other hand, our galaxies have lower (but still high) SFRs and such $\sim\!10^{12}\,\Msun$ haloes start to depart from the typical node at these redshifts, therefore not being replenished as efficiently with cold gas as similar haloes at higher $z$. We also add below each panel the direct comparison with NIHAO values.

Finally, we summarise the median values of each sample in Table~\ref{tab:mdot_gas}. Rates are shown at \Rgal, \Rvir, and $3\,\Rvir$ and show the corresponding cold fraction as well. Note that the net accretion is computed for each individual galaxy. Since the median is a statistical measure that does not preserve additive properties, the median net accretion rates are not expected to match the difference between the median inflow rates and the median outflow rates.
\begin{table}
    \centering
    \begin{tabular}{lcccc}
        \hline
        Type & $z$ & $\Rgal$ & $\Rvir$ & $3\,\Rvir$\\
        \hline
        In & 3.6 & 160 & 240 & 340\\
         & 2.0 & 67 & 110 & 12\\
         & 0.0 & 7 & 15 & 12\\
        Out & 3.6 & 80 & 115 & 0\\
        & 2.0 & 35 & 60 & 10\\
        & 0.0 & 6 & 11 & 8\\
        Net & 3.6 & 90 & 130 & 340\\
        & 2.0 & 25 & 20 & 150\\
        & 0.0 & -1 & 7 & -1\\
        \hline
        Cold (in) & 3.6 & 135 & 105 & 290\\
        & 2.0 & 50 & 33 & 113\\
        & 0.0 & 5 & 7 & 9\\
        Hot (in) & 3.6 & 30 & 120 & 70\\
        & 2.0 & 16 & 72 & 47\\
        & 0.0 & 3 & 7 & 3\\
        \hline
        $\fcold$ (in) & 3.6 & 0.80 & 0.46 & 0.80\\
        & 2.0 & 0.75 & 0.31 & 0.70\\
        & 0.0 & 0.65 & 0.46 & 0.75\\
        \hline
    \end{tabular}
    \caption{Median values of inflow, outflow and net accretion rates (upper set), cold and hot accretion rates (middle set), and the corresponding \fcold (lower set). The values, referring to our three samples at $z = 3.6$, 2.0, and 0.0, are estimated across different radii, i.e. \Rgal, \Rvir, 3\,\Rvir. All the rates are in units of $\Msun\,\mathrm{yr}^{-1}$.}
    \label{tab:mdot_gas}
\end{table}

\section{Cold fractions at final redshifts}\label{sec:cold_fraction_final_z}
We now investigate the accretion modes of our galaxies with our fiducial method defined in \S\ref{sec:cold_accretion}. That is, we do not resort to the instantaneous state of gas particles but also take into account their historical states by tracking their temperature across all snapshots and using their maximum temperature across all epochs to classify them as hot or cold.

\begin{figure*}
\centering
\includegraphics[width=\linewidth]{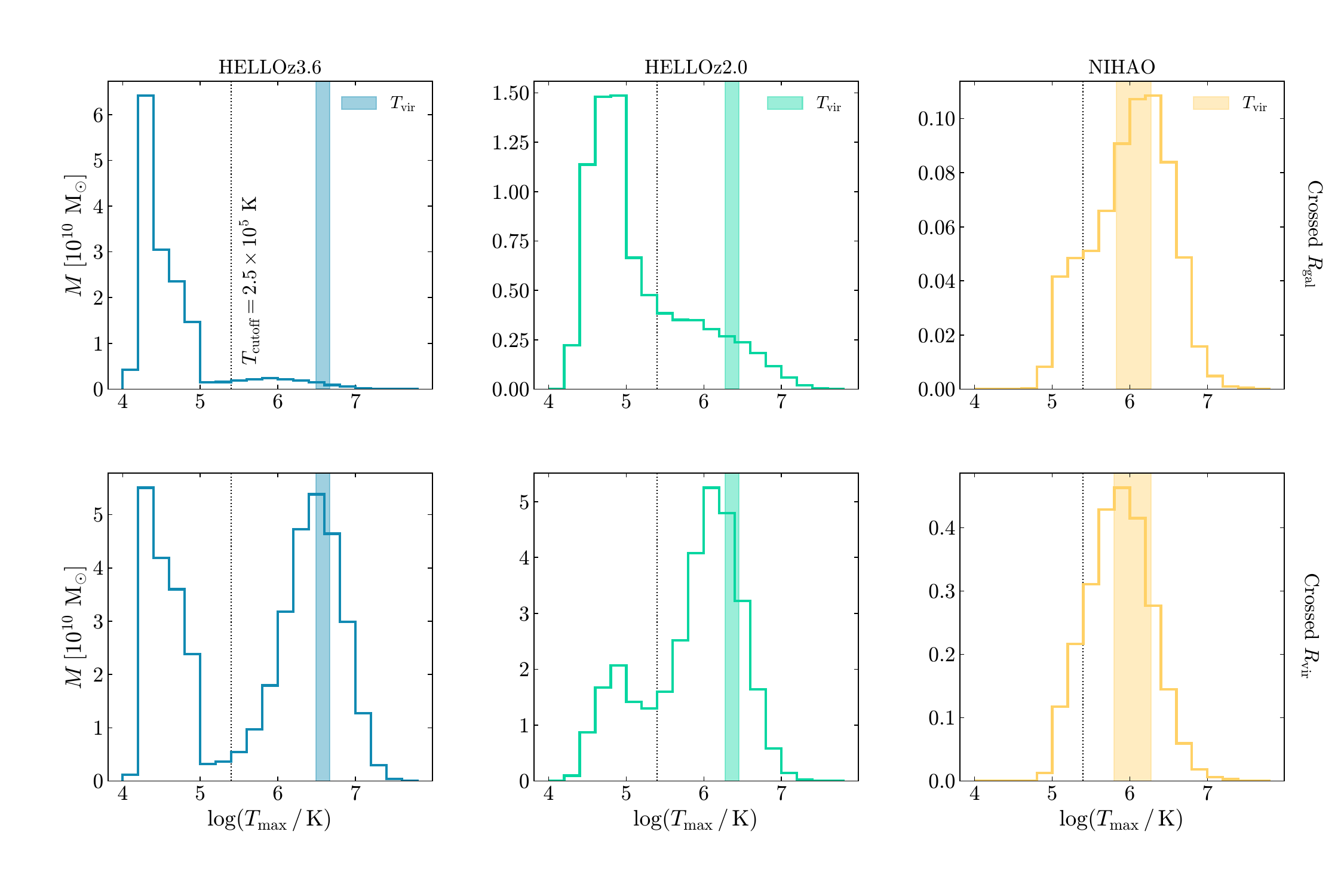}
\caption{\Tmax\ distribution for particles that crossed \Rgal\ (top) and \Rvir\ (bottom) in the last snapshot. Each histogram represents the combination of the \Tmax\ distributions across all galaxies in their respective samples, weighted by the particle mass. \Tmax\ is defined as the historical maximum temperature for each particle. For particles crossing \Rgal, \Tmax\ is computed using all \emph{previous} snapshots, while for particles crossing \Rvir, it is calculated across \emph{all} snapshots, including the last one. The histograms are constructed in logarithmic temperature bins from 4 to 8, with a bin width of 0.2. Galaxies with fewer than 100 particles crossing the respective radius are excluded from the analysis. From left to right, the panels show results for the HELLOz3.6, HELLOz2.0, and NIHAO samples, respectively.}
\label{fig:temp_max_hist}
\end{figure*}
We begin by plotting the distribution of \Tmax\ of the smooth gas that crossed \Rgal\ (top) and \Rvir\ (bottom) in Fig.~\ref{fig:temp_max_hist}. \Tmax\ is simply the maximum temperature that the particles had at any point in history, with the exception of the last snapshot for particles that crossed \Rgal. The distributions of all galaxies within each sample are combined into a single distribution, starting with HELLOz3.6 on the right, HELLOz2.0 in the middle, and NIHAO on the right, and weighted by the mass of each particle. The histograms are constructed in logarithmic temperature bins from 4 to 8, with a bin width of 0.2. We also display the virial temperatures of our different haloes as a band covering [$\mathrm{min}\,\Tvir$,~$\mathrm{max}\,\Tvir$], and the temperature cutoff $\Tcutoff = 2.5 \times 10^{5}$ for distinguishing cold and hot accretion as a vertical dotted line. Note that the galaxies for which there are less than 100 particles that crossed the radius of interest are not included in the histograms.

HELLO galaxies overall exhibit a bimodal distribution for both galaxy and halo accretion. NIHAO galaxies do not present a significant bimodality, which completely vanishes for the smooth accretion onto the halo. Virtually all particles accreted onto the galaxy and the halo at $z=0$ reached temperatures of at least $10^5$ K at some point, in stark contrast to HELLO galaxies for which a significant number of particles remain below $10^5$ K at all times.

\begin{figure*}
\centering
\includegraphics[width=\linewidth]{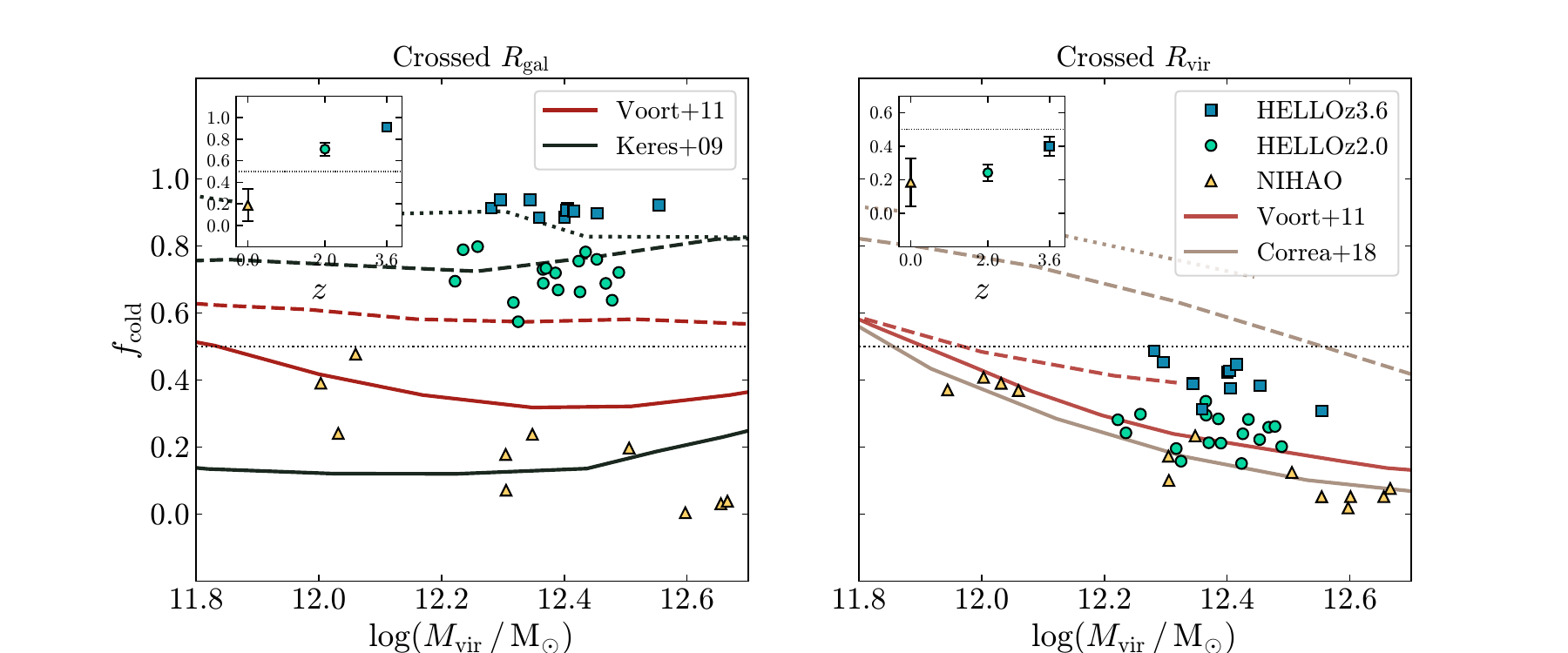}
\caption{Fraction $f_{\mathrm{cold}}$ of gas accreted onto the galaxy (left) and the halo (right) whose temperature never exceeded \Tcutoff, plotted against the virial mass of the galaxy. Each marker represents \fcold of one galaxy, colour-coded by the parent sample (HELLOz3.6, blue squares; HELLOz2.0, turquoise circles; and NIHAO, yellow triangles). The dotted line shows where $\fcold = 0.5$, above which galaxies are in the cold-dominated regime. The insets in both panels show the average and standard deviation \fcold of each sample at their respective redshift. Galaxies for which there are less than 100 particles that crossed the radius of interest are not included in the figure. Overplotted are previous results from \citet{Keres2009a} (black), \citet{Voort2011a} (red) and \citet{Correa2018a} (grey). Line styles follow the same convention as Fig.~\ref{fig:mdot_comparison}, i.e. solid for $z=0$, dashed for $z=2$ and dotted for $z=4$.}
\label{fig:cold_frac_nihao_vs_hello}
\end{figure*}
Moving on to Fig.~\ref{fig:cold_frac_nihao_vs_hello}, we show the fraction of gas that has entered the galaxy (left) and the halo (right) between the last snapshot and the one before that never exceeded \Tcutoff, plotted against the virial mass at the time of crossing. The galaxies are colour-coded with respect to the sample they belong to, with HELLOz3.6 in blue squares, HELLOz2.0 in turquoise circles, and NIHAO in yellow triangles. The average of each sample as a function of redshift is shown in the inset on the upper left of each panel and can be found in Table~\ref{tab:cold_fractions} as well.
\begin{table}
    \centering
    \begin{tabular}{cccccc}
        \hline
        Suite & $z$ & Accretion onto & $f_{\mathrm{cold}}$\\
        \hline
        HELLO & 3.6 & Galaxy & 91\%\\
        HELLO & 3.6 & Halo & 40\%\\
        \hline
        HELLO & 2.0 & Galaxy & 71\%\\
        HELLO & 2.0 & Halo & 24\%\\
        \hline
        NIHAO & 0.0 & Galaxy & 19\%\\
        NIHAO & 0.0 & Halo & 19\%\\
        \hline
    \end{tabular}
    \caption{Average cold fraction of gas particles accreted onto the the galaxy and the halo for HELLOz3.6, HELLOz2.0 and NIHAO galaxies.}
    \label{tab:cold_fractions}
\end{table}

Beginning with the left panel, the main feature is the significant redshift dependence on \fcold, as highlighted by the plot inset. More than 90~per~cent of gas particles entering the galaxy at $z=3.6$ have always been cold and this number reduces to roughly 70~per~cent at $z=2$. Our high-$z$ galaxies are thus well into the cold-dominated regime with respect to the particles that were just accreted onto the galaxy. NIHAO simulations, on the other hand, are all below $\fcold = 0.5$, varying from about 40 per cent at $\approx10^{12}\,\Msun$ down to just a few percentage points at $4\times10^{12}\,\Msun$, implying an almost negligible fraction of cold gas, and therefore hot dominated.

While less prominent, the redshift dependence still appears on the right panel between HELLOz3.6 and HELLOz2.0, now showing the gas particles that crossed \Rvir\ during the same interval as before. NIHAO galaxies do not display different results from the left panel and show a similar $\fcold \sim 0.2$ on average. HELLO galaxies, however, are all hot dominated. Galaxies at $z=3.6$ still accrete a rather large fraction of cold gas ($\sim\!40$~per~cent) but the majority of the gas is nonetheless accreted hot.

The higher fraction of cold accretion at \Rgal\ at high-$z$ suggests that the cold accretion is mainly stemming from cold streams reaching the galaxy. Indeed, while accretion onto the halo shows a significant hot accretion, these particles do not appear to reach the galaxy and likely remain in the hot CGM where the cooling time exceeds the local dynamical time and we note that these haloes have masses consistent with the formation of a hot halo atmosphere in hydrostatic equilibrium \citep[e.g.][]{Correa2018a}.

Regarding the hot accretion onto the halo, it is also interesting to investigate how many particles were heated inside the halo after crossing \Rvir\ in comparison to particles that were `truly' accreted hot, namely those heated before crossing \Rvir. We obtain the following median fractions of hot accreted gas heated inside \Rvir: $0.24^{+0.17}_{-0.09}$, $0.10^{+0.04}_{-0.05}$, and $0.03^{+0.01}_{-0.02}$ for HELLOz3.6, HELLOz2.0, and NIHAO, respectively, with the uncertainties covering the 16th to 84th percentile range. These results show that in all cases, the majority of the hot accreted gas entered the halo already hot and are consistent with Figs.~\ref{fig:temp_tracks_vs_time} and \ref{fig:large_scale_maps}.

As before, we compare our galaxies with previous results using a similar method to define cold accretion, i.e. by tracking the historical temperature state of the gas particles in this case. Starting with \citet{Voort2011a}, both panels show their results in red at $z=0$ (solid lines) and $z=2$ (dashed lines). Except for accretion onto \Rgal\ at $z=2$, our galaxies are generally biased towards lower cold fractions. Their method to quantify the final cold fraction is similar to ours, although with slight variations. Their galaxy radius is defined as $0.15\,\Rvir$, smaller than our definition of $0.2\,\Rvir$. This could in theory allow some gas particles closer to the galaxy and therefore increase their chance of interacting with the hot ISM, leading to lower cold fractions, especially at higher redshift where star formation is highly active. At $z=0$, the tension in the left panel mainly stems from the different trends observed. NIHAO galaxies show a clear linear dependence with mass, while galaxies from \citet{Voort2011a} exhibit a shift towards increasing \fcold\ above $\log(\Mvir/\Msun) \sim 12.5$. Overall, however, both our galaxies and theirs exhibit cold (hot) mode-dominated accretion at $z=2$ ($z=0$) and strong redshift dependence.

Also shown on the same panel are the cold fractions found by \citet{Keres2009a} (black) at $z=0$ (continuous), $z=2$ (dashed), and $z=4$ (dotted). HELLO simulations are in excellent agreement at high-$z$, while at $z=0$ their data exhibits a similar trend than \citet{Voort2011a} towards high masses. In their analysis, \citet{Keres2009a} attribute the rise of the cold fraction at low redshift and high masses to a combination of vanishing hot mode accretion rates and `cold drizzle'. They identify the latter as cold clouds stemming either from the remnants of cold filaments, either from clouds not identified as galaxies (thus counted as smooth accretion), or cold galactic gas stripped away by ram pressure and tidal forces. They note that most of this cold drizzle is likely a numerical artefact and also depends on the code used, with e.g. \textsc{\small GASOLINE} resulting in lower amounts of cold drizzle. Moreover, given our conservative definition of smooth accretion, any particle reaching the galaxy at $z=0$ either entered the halo `recently', in which case the probability of it being shock heated at these halo masses is high, or it entered the halo for the first time at much earlier times, implying that it was recycled or cooled from the hot halo, both cases prohibiting the particle to be counted as cold.

\begin{figure}
\centering
\includegraphics[width=\columnwidth]{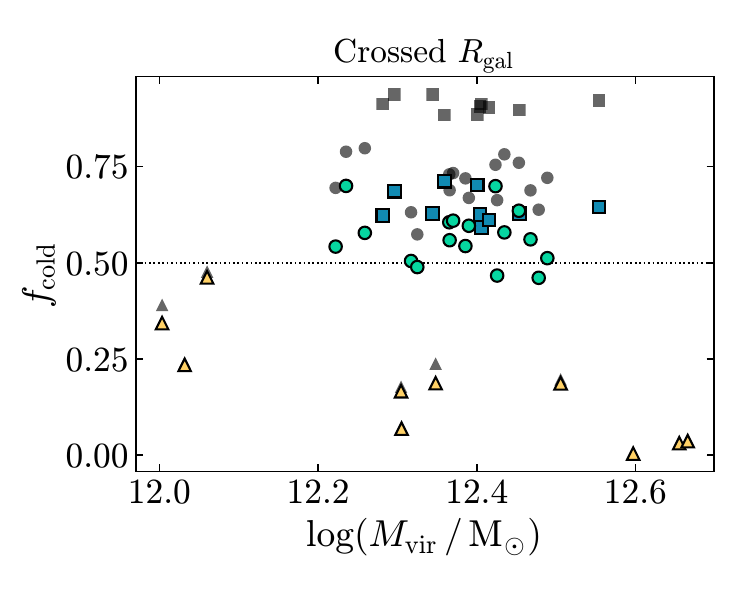}
\caption{Same as the left panel of Fig.~\ref{fig:cold_frac_nihao_vs_hello} where the coloured markers now represent the fraction $f_{\mathrm{cold}}$ of gas that was accreted onto the galaxy, when accounting for the last snapshot in the temperature history. The grey markers represent the data points from Fig.~\ref{fig:cold_frac_nihao_vs_hello}.}
\label{fig:cold_frac_ism}
\end{figure}

Before describing the cold fraction for particles accreted onto the halo in the right panel, we want to emphasise that while the exact definition of cold accretion can significantly alter the results, this is even more true for the accretion onto the galaxy. The closer one is to the central galaxy, the higher the chance is that a gas particle gets directly impacted by stellar and AGN feedback. Therefore, variations in sub-grid models as well as in the choice of \Rgal\ make fair comparisons across simulations difficult. We illustrate this by recalculating the cold fractions in our simulations when considering the last snapshot, i.e. when the galaxy is within \Rgal, as part of the temperature history. The results can be seen in Fig.~\ref{fig:cold_frac_ism}, where the squares represent the updated values (the colour scheme is kept as before) and the grey circles are the same as in Fig.~\ref{fig:cold_frac_nihao_vs_hello}. The cold fraction of HELLO galaxies is now lower, with a significant drop from $\sim\!90$~per~cent to $\sim\!60$~per~cent for HELLOz3.6. Due to the high SFRs at high-$z$ and the subsequent large feedback energy released, \fcold\ in massive star-forming galaxies is hence highly sensitive to where it is computed. Unsurprisingly for their quenched counterparts at $z=0$, such sensitivity vanishes.

Regarding accretion onto the halo, on top of \citet{Voort2011a} we also compare our galaxies to \citet{Correa2018a}, both of which use SPH codes. At $z=0$, NIHAO galaxies are in excellent agreement with both works. While our galaxies are again slightly biased towards lower \fcold\ at high $z$ compared to \citet{Voort2011a}, the disagreement is quite significant compared to \citet{Correa2018a}. The latter does not use the historical maximum gas temperature \Tmax\ but rather the gas temperature right after accretion. They also do not discriminate between smooth and merger accretion but find that it mostly affects the gas accretion rate rather than the cold fraction itself. We will discuss their claim in more detail in the next section. 

\section{The cosmic evolution of cold gas fraction}\label{sec:fcold_cosmic_evolution}
In this section, we explore how \fcold\ of individual galaxies evolves with cosmic time.
\begin{figure*}
\centering
\includegraphics[width=\linewidth]{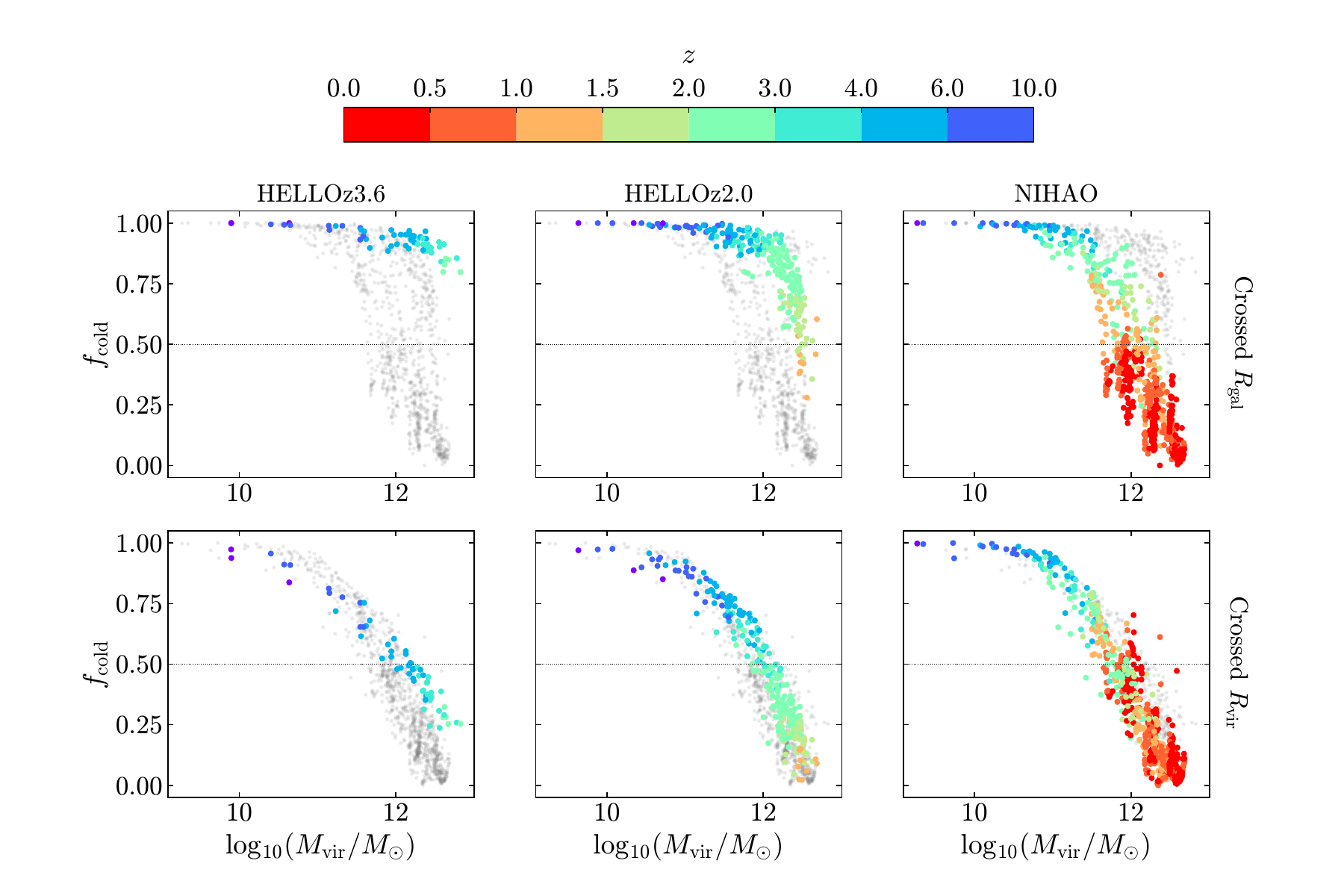}
\caption{Cold fraction evolutionary tracks from particles crossing the galaxy (top) and the halo (bottom) for HELLOz3.6 (left), HELLOz2.0 (middle), and NIHAO (right) colour-coded by redshift. Additionally, each panel contains the other two samples in grey. Dots represent all snapshots where there were at least 100 particles crossing the relevant boundary.}
\label{fig:cold_frac_evolution}
\end{figure*}
Fig.~\ref{fig:cold_frac_evolution} shows the evolutionary tracks of the cold gas fraction \fcold accreted onto galaxies (top) and halos (bottom), for HELLOz3.6 (left), HELLOz2.0 (middle), and NIHAO (right). Each panel includes all snapshots (grey dots) for which \fcold could be defined, i.e. where there are \emph{at least} 100 gas particles crossing the galaxy/halo and where the parent halo was present in the preceding snapshot.
Overall, we retrieve the observations we made in the previous section: for accretion onto galaxies, Fig.~\ref{fig:cold_frac_evolution} exhibits a clear redshift dependence, especially for HELLOz3.6 that seems to still be in a relatively flat phase where \fcold has not yet started to decline sharply. As a consequence of this evolution, the cold-to-hot phase transition occurs at higher critical mass with increasing redshift.

Concerning the evolution of cold fraction crossing the halo, we observe a global tighter distribution. However, as we will demonstrate in \S\ref{subsec:bayesian_model}, a redshift dependence is found in particular at $z\gtrsim1$.

We already pointed out earlier that \fcold for particles accreted onto the galaxy can significantly vary depending on the definition adopted, especially at high redshift, where high SFRs lead to a large amount of energy being released into the ISM. We therefore opt to focus our analysis on the accretion onto the halo, as it has been shown \citep[e.g.,][]{Voort2011a, Voort2011b, Nelson2015} that feedback, whether from stars or AGN, is expected to have a lower impact for accretion at \Rvir.
Additionally, we qualitatively test for different definitions of cold accretion and find that the accretion onto the halo in our simulations is more robust to the various methods, see Appendix~\ref{app:cold_frac_evol}, where we show the evolutionary tracks for our different tests. In the remaining section, we focus on \fcold\ at \Rvir, determined from our fiducial method (smooth gas, Lagrangian, and \Tmax\ criterion).

\subsection{Modelling the cold fraction evolution with time and mass}\label{subsec:bayesian_model}
Our goal is to propose a continuous model for describing the redshift evolution of \fcold\ as a function of the halo mass, i.e. find a relation describing $\fcold \equiv \fcold(\Mvir,z)$. In \citet{Correa2018a}, the authors developed a model which leverages least-square minimisation at individual redshifts, then fitting the redshift evolution of the model parameters. In this work, instead, we exploit a Bayesian hierarchical framework providing a continuous description of the back-in-time evolution of \fcold\ combining the data from the NIHAO, HELLOz2.0 and HELLOz3.6 samples.
\subsubsection{The model}\label{subsubsec:model}
To reconstruct the evolution of \fcold, both the virial mass \Mvir\ and the redshift $z$ of each simulated object are considered as independent variables, while the cold gas fraction \fcold\ represents the dependent variable of our model.
As clearly visible in Fig.~\ref{fig:cold_frac_evolution}, data points are distributed on the $\Mvir{-}\fcold$ plane as a sigmoid function where the transition critical mass between cold and hot dominated accretions is defined as the mass where $\fcold=0.5$. The common picture from theory \citep[e.g.,][]{Dekel2006} is that two different regimes in redshift can be identified:
\begin{itemize}
    \item a \emph{low-z} regime, which presents an almost constant critical mass with redshift (\Mshock);
    \item a \emph{high-z} regime, in which the critical mass increases towards high redshifts (\Mstream).    
\end{itemize}
This transition between the low- and high-$z$ regimes is found to occur typically at $z_\tau\approx1{-}2$. The presence of these two regimes motivates our choice to model the evolution of \fcold\ adopting a \emph{double sigmoid function}:
\begin{equation}\label{eq:model}
    f_\mathrm{cold}(\Mvir,z,\Hyper)=\mathcal{S}_1\,\tau+
    \mathcal{S}_2\,(1-\tau),
\end{equation}
where $\Hyper$ is the set of hyperparameters of our model (refer to Appendix \ref{app:bayesian}), while $\mathcal{S}_i$ (with $i \in \{1,2\}$) refers to a sigmoid function of the form
\begin{equation}\label{eq:sigmoid}
    \mathcal{S}_i\equiv\mathcal{S}_i(x,\kappa_{i},x_{\mathrm{c},i}) = \frac{1}{1 + e^{-\kappa_{i}(x - x_{\mathrm{c},i})}}.
\end{equation}
In Eq.~\ref{eq:model}, $\mathcal{S}_1\equiv\mathcal{S}_1(\log \Mvir,\kappa_{1},\log M_{c,1})$ models the evolution of \fcold at $z\lesssim z_\tau$, while $\mathcal{S}_2\equiv\mathcal{S}_2(\log \Mvir,\kappa_{2},\log M_{c,2})$ at $z\gtrsim z_\tau$. The coefficients $\kappa_{i}$ and $x_{c,i}$ in Eq.~\ref{eq:sigmoid} represent the sigmoid's steepness and the inflexion point, respectively. In our Bayesian approach, we let both coefficients evolve with redshift as
\begin{equation}\label{eq:steepness}
    \kappa_{i}\equiv\kappa_{i}(z)=\kappa_{0,i}+\kappa_{z,i}\log(1+z),
\end{equation}
and
\begin{equation}\label{eq:inflexion}
    x_{\mathrm{c},i}\equiv x_{\mathrm{c},i}(z)=x_{0,i}+x_{z,i}\log(1+z),
\end{equation}
with $\kappa_{0,i}$, $\kappa_{z,i}$, $x_{0,i}$, and $x_{z,i}$ as hyperparameters of our framework.

The coefficient $\tau$ in Eq.~\ref{eq:model} models the transition between the two above-mentioned regimes. To model this transition in redshift, we adopt another sigmoid of the same form as Eq.~\ref{eq:sigmoid}, i.e.
\begin{equation}\label{eq:transition}
    \tau \equiv \mathcal{S}_\tau(z,\kappa_{\tau},z_\tau),
\end{equation}
with $\kappa_{\tau}$ and $z_\tau$ as two hyperparameters of the model.

In Appendix~\ref{app:bayesian} we describe in detail the Bayesian hierarchical approach adopted to trace the back-in-time evolution of the $\fcold{-}\Mvir$ relation. Moreover, in Tab.~\ref{tab:hyperparameters}, we provide a description of the hyperparameters in our model including the priors adopted.

\subsubsection{Defining an independent dataset in redshift}\label{ssec:independent_data}
A source of bias in our simulated samples to account for is the non-independent nature of data. Indeed, our data are drawn from multiple snapshots, meaning that each galaxy is present at different stages of its evolution. To reduce this effect in our Bayesian analysis, we take advantage of the fact that the evolutionary track of each galaxy can be considered as a time series. This implies that we can calculate the autocorrelation function (ACF) and find the corresponding lag (i.e. separation) at which two data points can be considered statistically independent of each other. This procedure is detailed in Appendix~\ref{app:indep_data} and the final sample consists of 246 independent points.

\subsection{Results}
\label{ssec:results}
In the following, we present the results of the analysis aimed at tracing the back-in-time evolution of the cold gas fraction over cosmic time. The posterior probability distribution of our model is sampled exploiting \textsc{dynesty} \citep{Speagle2020MNRAS,Koposov2023zndo}, a pure \textsc{python} package implementing the \emph{Nested Sampling}
technique \citep{Skilling2004AIPC,Skilling2006}.
In Table~\ref{tab:dynesty_table}, we list all the parameters used to set up our runs.
\begin{table}
    \centering
    \begin{tabular}{lc}
        \hline
        Parameter                & Value  \\
        \hline
        {\ttfamily nlive      }  & 2000   \\
        {\ttfamily bound      }  & multi  \\
        {\ttfamily sample     }  & rslice \\
        {\ttfamily queue\_size}  & 8        \\
        {\ttfamily dlogz}        & 0.01   \\
        {\ttfamily rstate}       & numpy.random.default\_rng(18)\\
        \hline
    \end{tabular}
    \caption{List of settings used for our \textsc{dynesty} run. Column 1: parameter of the {\ttfamily NestedSampler} class and the {\ttfamily run\_nested} method. Column 2: parameter values.}
    \label{tab:dynesty_table}
\end{table}

In Table~\ref{tab:values}, we list the medians and the associated $1\sigma$ uncertainties of the hyperparameters sampled for our model.
The evolution of the terms $\kappa_i$ and $\log M_{c,i}$ (refer to Eq.~\ref{eq:inflexion} and Eq.~\ref{eq:steepness}) as well as of the transition term $\tau$ (see Eq.~\ref{eq:transition}) are shown in Fig.~\ref{fig:hparams_evo} of the corresponding appendix.
\begin{table}
\begin{center}
\renewcommand{\arraystretch}{1.5}
\begin{tabular}{cc} 
 \hline
 Hyperparameter & Value\\
 \hline
 $\kappa_{0,1}$   & $-3.60_{-0.67}^{+0.66}$ \\
 $\kappa_{z,1}$   & $2.40_{-4.20}^{+3.00}$ \\
 $\log M_{0,1}$   & $11.83_{-0.09}^{+0.06}$ \\
 $\log M_{z,1}$   & $-0.18_{-0.36}^{+0.60}$ \\
 $\kappa_{0,2}$   & $-2.65_{-0.19}^{+0.20}$ \\
 $\kappa_{z,2}$   & $1.50_{-0.25}^{+0.23}$ \\
 $\log M_{0,2}$   & $10.94_{-0.12}^{+0.11}$ \\
 $\log M_{z,2}$   & $1.69_{-0.18}^{+0.19}$ \\
 $z_\tau$         & $1.23_{-0.29}^{+0.38}$ \\
 $\sigma_0$       & $0.08_{-0.01}^{+0.01}$ \\
 $\sigma_z$       & $-0.04_{-0.02}^{+0.02}$\\
 \hline
\end{tabular}
\caption{Inferred medians and 68 per cent posterior credible ranges of the hyperparameters.
         Column 1: name of the hyperparameter.
         Column 2: median values with the 16-th (lower value) and 84-th (upper value) percentile uncertainties.}
\label{tab:values}
\end{center}
\end{table}
Using the values listed in Table~\ref{tab:values} coupled to Eqs.~\ref{eq:model}--\ref{eq:transition}, the evolution of \fcold\ as a function of \Mvir\ and $z$ can be approximately described by
\begin{equation}
    f_{\mathrm{cold}}(\Mvir,z) = \frac{\tau}{1+e^{-\kappa_1(\log\Mvir-\log M_{\mathrm{c},1})}}
    +\frac{1-\tau}{1+e^{-\kappa_2(\log\Mvir-\log M_{\mathrm{c},2})}},
\end{equation}
with
\begin{equation}
\begin{aligned}
    &\kappa_1 \simeq\, -3.60+2.40\log(1+z),\\
    &\kappa_2 \simeq\, -2.65 + 1.50\log(1+z),\\
    &\log M_{\mathrm{c},1} \simeq\, 11.83 - 0.18\log(1+z),\\
    &\log M_{\mathrm{c},2} \simeq\, 10.94 + 1.69\log(1+z),\\
    &\tau \simeq\, \frac{1}{1+e^{10(z-1.23)}}.
\end{aligned}
\end{equation}
We do find evidence for a weak evolution of the intrinsic scatter of the relation that can be roughly described by
\begin{equation}
    \sigma_{\fcold} \simeq 0.08 -0.04\log(1+z),
\end{equation}
\begin{figure*}
\centering
\includegraphics[width=\linewidth]{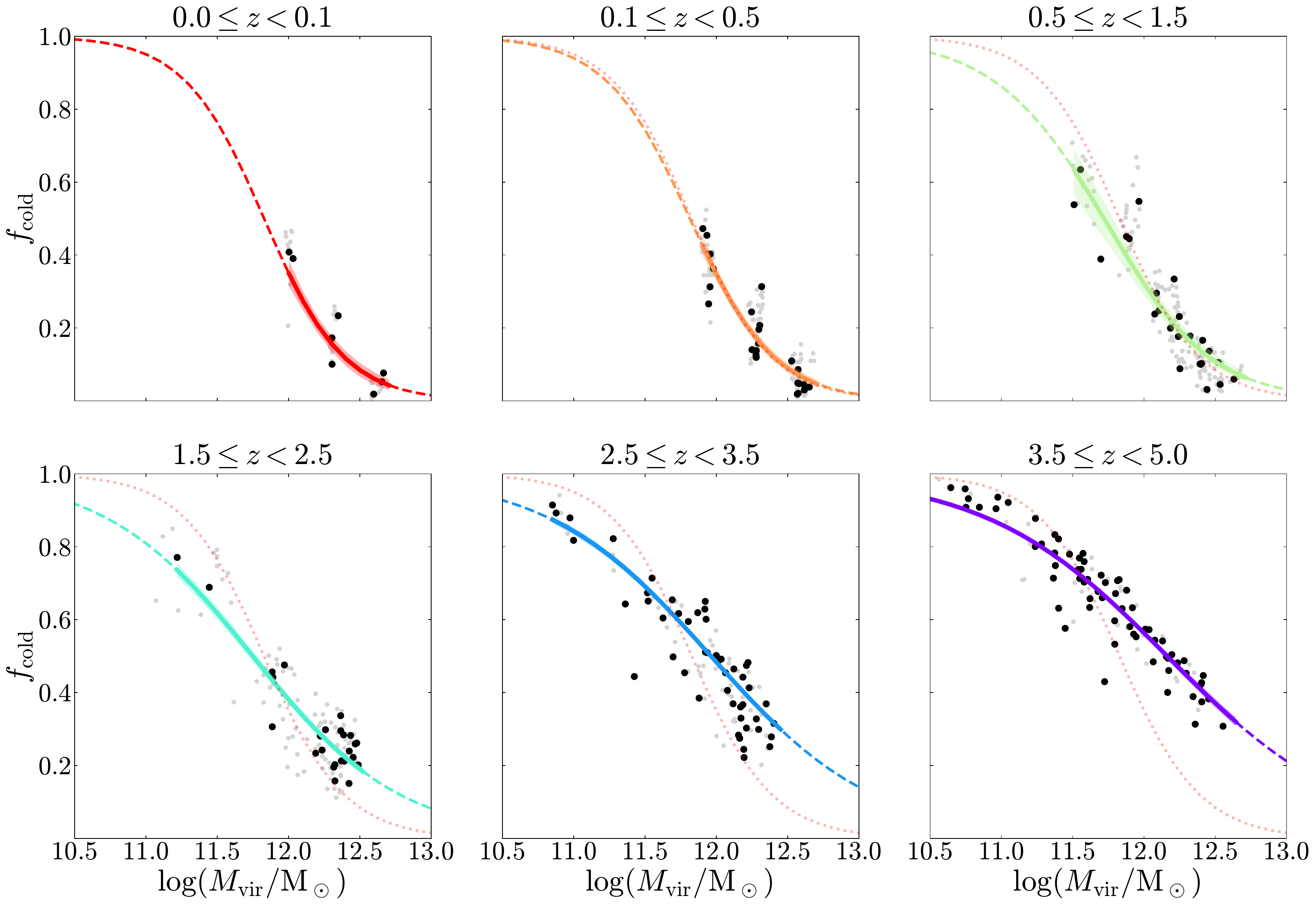}
\caption{Evolution of \fcold\ as a function of \Mvir in six different redshift bins. Each panel contains the model evaluated at the redshift bin's midpoint, displayed as a dashed line and the 1$\sigma$ uncertainty region, delimited by the 16~per~cent and the 84~per~cent of the distribution of \fcold at fixed \Mvir. The continuous curves represent the mass range covered by the data classified as independent (shown as larger black dots). Shaded grey dots include all snapshots in the respective bin. The red-shaded dotted curve in each panel corresponds to the model evaluated at $z=0$ (top left panel).}
\label{fig:fcold_bins}
\end{figure*}
The median $\fcold{-}\Mvir$ relation evaluated at the midpoints of six redshift bins is shown in Fig.~\ref{fig:fcold_bins}.
Additionally, the black circles in each panel represent all the snapshots classified as independent (see Appendix~\ref{app:indep_data}) in the corresponding redshift range that were fed to the algorithm, while the shaded circles represent all snapshots in that bin. Each curve is extrapolated beyond the mass range covered by the data, shown as dashed extensions. We also add in each subplot the \fcold\ relation at $z=0$, represented as a shaded dotted curve, to highlight the transition between constant \Mshock\ at low-$z$ and evolving \Mstream\ at high-$z$.

\begin{figure}
\centering
\includegraphics[width=\linewidth]{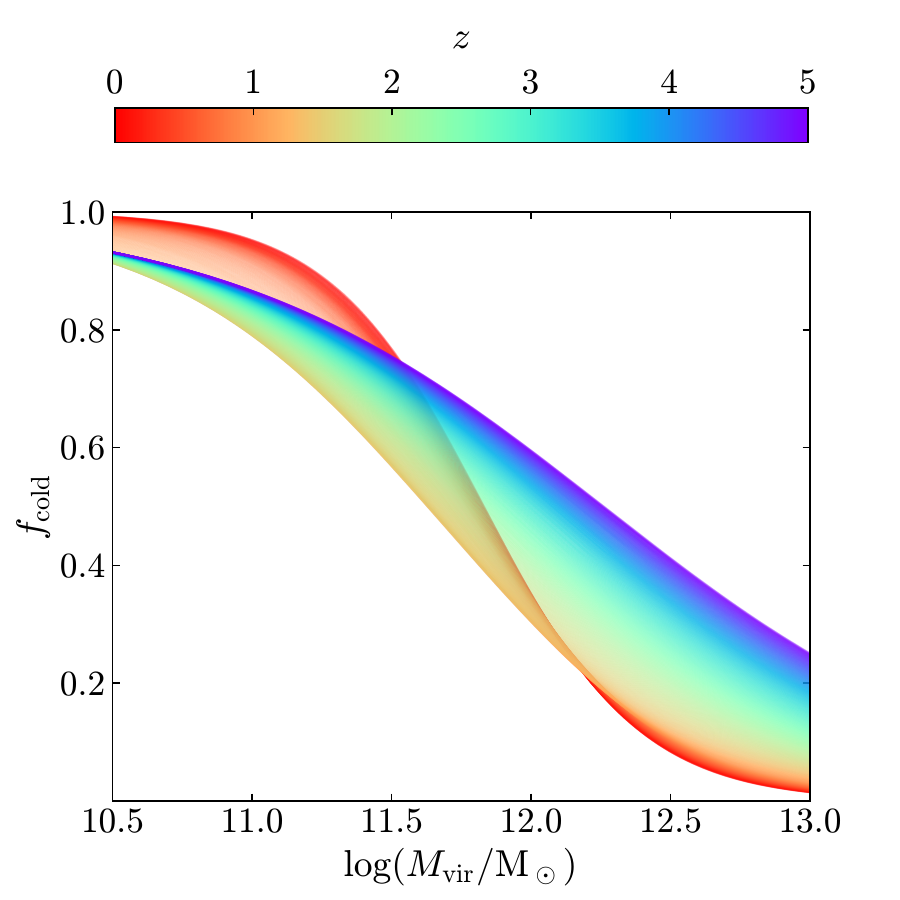}
\caption{Visual representation of the continuous evolution of our $\fcold{-}\Mvir$\ relation for $\log(\Mvir/\Msun) \in [10.5,13.0]$ and $z \in [0,5]$.}
\label{fig:fcold_rainbow}
\end{figure}

Each curve recovers the underlying snapshots, both independent and all, rather well. This supports the ability of our model to reproduce the population \fcold\ evolution of HELLO galaxies. As expected, \fcold\ decreases with mass at fixed $z$ and \fcold\ for massive galaxies increases with redshift. Finally, a visual representation of the continuous redshift evolution of the relation over the virial mass range $\log(\Mvir/\Msun) \in [10.5,13]$ and redshift range $z\in[0,5]$ is shown in Fig.~\ref{fig:fcold_rainbow}.

\subsection{Comparison with previous work}
In the final part of our analysis, we compare the \fcold\ evolution inferred from our model with previous results from \citet{Ocvirk2008, Voort2011a}, and \citet{Correa2018a}. In addition, we compare the evolution of the \Mshock\ and \Mstream\ with theoretical predictions from \citet{Dekel2006}, simulations from \citet{Ocvirk2008}, and recent observations from \citet{Daddi2022b}. The main differences between the methods to infer \fcold\ from previous numerical studies and this one are summarised in Table~\ref{tab:literature_methods}.
\begin{table*}
\begin{center}
    \begin{tabular}{lcclccccc} 
 \hline
 Paper & Hydro & Feedback & Acc. method & Acc. type & Criterion & Accretion through\\
 \hline
 \citet{Ocvirk2008} & AMR & SN & Eulerian & Smooth & $T_{\mathrm{gas}}$ & [0.2-1]$\,\Rvir$\\
 \citet{Voort2011a} & SPH & SN & Lagrangian & Smooth & \Tmax & \Rvir\\
 \citet{Correa2018a} & SPH & SN+AGN & Lagrangian & All & $T_{\mathrm{gas}}$ & \Rvir\\
 \textbf{This work} & SPH & ESF+SN+AGN & Lagrangian & Smooth & \Tmax & \Rvir\\
\hline
\end{tabular}
\caption{Summary of the main differences between previous works in the literature we compare our model to. The columns from left to right are the reference work, the type of code used, the feedback models implemented, the method to calculate gas accretion, the type of accretion, the temperature criterion used, and the definition of the accretion onto the halo. Regarding \citet{Voort2011a}, the authors also study the effect of AGN feedback, but we compare our results to their ‘REF’ model, which does not include AGN. All studies include metal-line cooling.}
\label{tab:literature_methods}
\end{center}
\end{table*}

\begin{figure*}
\centering
\includegraphics[width=\linewidth]{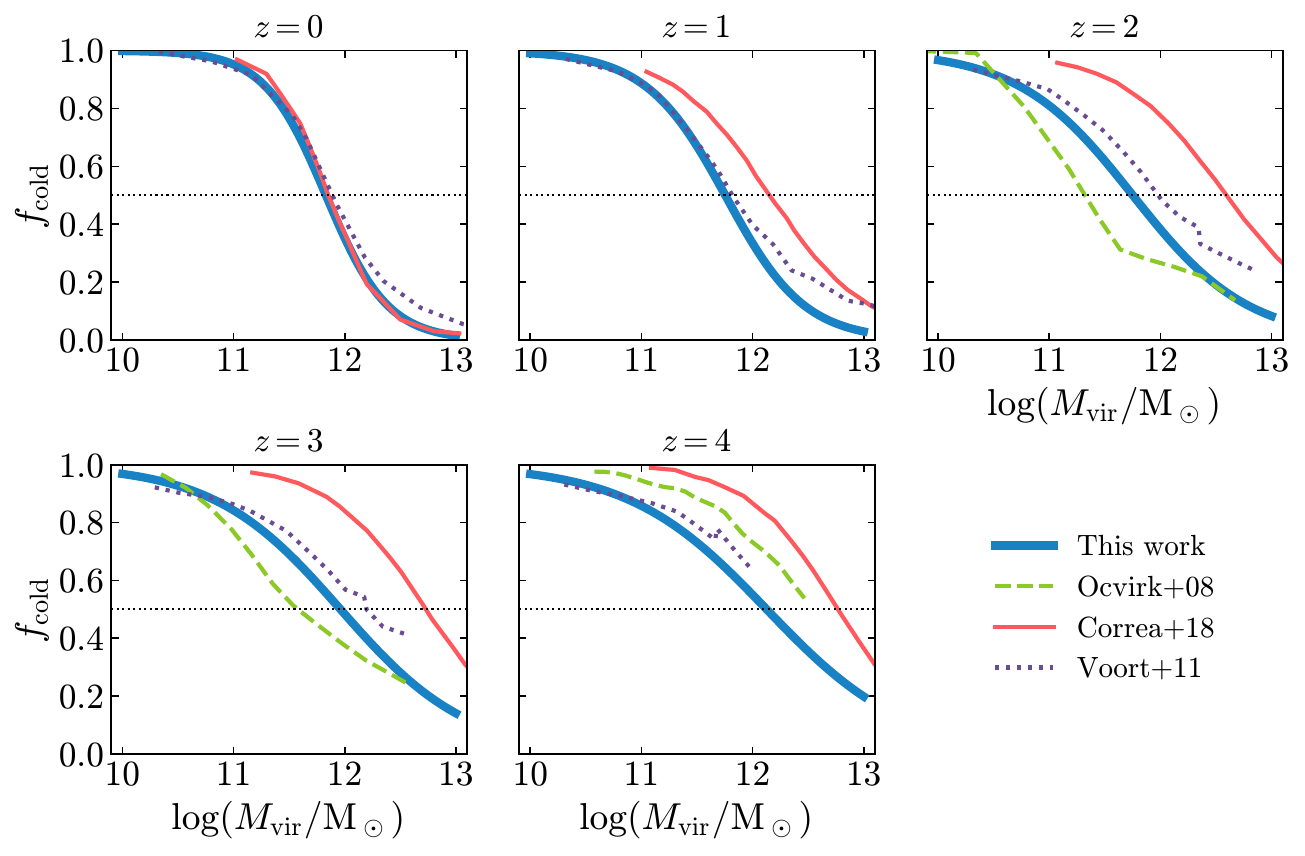}
\caption{Evolution of \fcold\ as a function of \Mvir\ from our model at fixed $z=\{0,1,2,3,4\}$ compared to previous works. Each panel contains the model evaluated at one redshift (continuous, blue) compared to results from \citet{Ocvirk2008} (dashed, green), \citet{Voort2011a} (dotted, purple), and \citet{Correa2018a} (continuous, orange). Apparent discontinuities in dotted curves arise because the data stems from two distinct simulation volumes at all redshifts (see their Fig.~11).}
\label{fig:model_comparison}
\end{figure*}
In Fig.\ref{fig:model_comparison}, we compare our model at $z=\{0,1,2,3,4\}$ with the measurements from \citet{Ocvirk2008, Voort2011a}, and \citet{Correa2018a}. Our model agrees remarkably well with \citet{Voort2011a} and \citet{Correa2018a} at $z=0$.\footnote{\citet{Ocvirk2008} do not compute the relation below $z=2$} At higher redshift, some tension appears with \citet{Ocvirk2008} and \citet{Correa2018a}.

\begin{figure}
\centering
\includegraphics[width=\columnwidth]{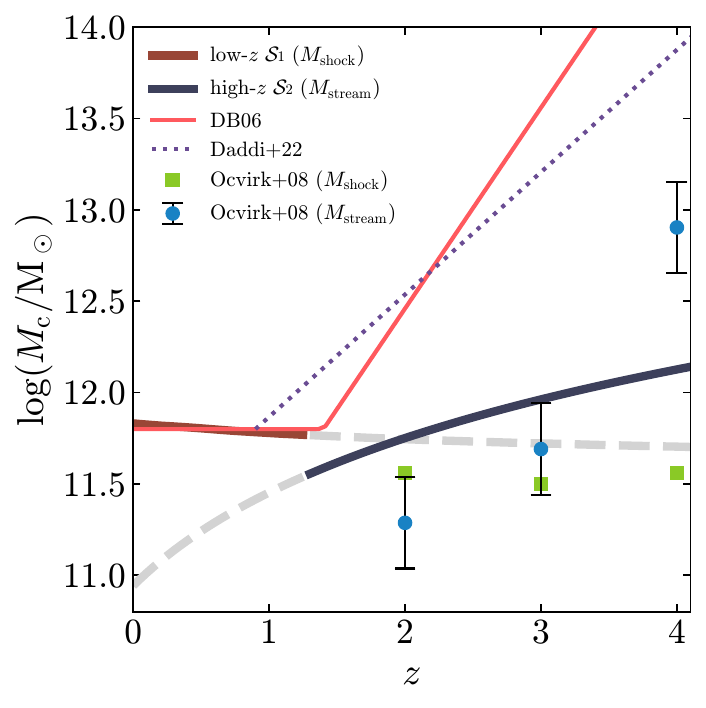}
\caption{Evolution of \Mshock\ and \Mstream\ inferred from our simulations compared with theoretical predictions \citep[][]{Dekel2006}, simulations \citep[][]{Ocvirk2008}, and recent observations \citep[][]{Daddi2022b}. The dashed line represents an extrapolation of \Mshock\ (\Mstream) above (below) the transition $z_\tau \simeq 1.26$. The transition in our model happens smoothly (not shown here) according to Eq.~\ref{eq:transition}. Note that \citet{Ocvirk2008} refers to \Mshock\ with respect to the accretion onto the \textit{galaxy}.}
\label{fig:mcrit_comparison}
\end{figure}

As already pointed out in \S\ref{sec:mass_accretion_rates}, \citet{Ocvirk2008} use an AMR code, which means that their method is Eulerian. Their simulations include weak SN feedback but no AGN feedback and their hot and cold accretions rates are integrated over the entire CGM. We predict a constant critical mass \Mshock\ up to $z \sim 1.2$ with a subsequent gradual increase of \Mstream\ with redshift (Fig.~\ref{fig:mcrit_comparison}), at odds with their relatively constant relation between $z=1-2$ and sudden change between $z=3-4$.

Unlike \citet{Ocvirk2008}, both \citet{Voort2011a} and \citet{Correa2018a} adopt the SPH technique in their simulations and define the accretion onto the halo similarly to this work: a gas particle has to be within \Rvir\ in the last snapshot but outside \Rvir\ in the snapshot before. However, \citet{Correa2018a} do not discriminate between smooth accretion and mergers and consider all gas particles. Moreover, unlike our method, they only measure the gas temperature in the final snapshot, i.e. \textit{after} entering the halo, to separate cold from hot inflows. We also previously pointed out that they claim that selecting all particles instead of smooth ones predominantly affects \textit{accretion rates} rather than \fcold.
While qualitative, Fig~\ref{fig:cold_frac_evolution_halo_comp} conveys a different interpretation from our simulations.
Fixing which particles are included (smooth or all; top versus bottom row), the temperature criterion (\Tmax or $T_{\mathrm{gas}}$; first versus second column), does not appear to significantly alter the overall distribution of \fcold\ values, except mainly at $z \lesssim 0.5$.
The latter is not surprising, as the temperature criterion is naturally expected to have a large impact if the gas has been going through multiple accretion episodes.
At high-$z$, on the other hand, Fig~\ref{fig:cold_frac_evolution_halo_comp} seems to be consistent with our interpretation of Fig.~\ref{fig:temp_max_hist} in the beginning of this study.
Unlike \citet{Correa2018a}, however, our figure suggests that the choice of which particles are included does have an impact on \fcold.
In summary, the disagreement between our model and \citet{Correa2018a} likely arises due to the combination of: i) different sub-grid physics, i) different choices of particles used, and iii) different temperature criteria used.

Finally, our model is overall very consistent with \citet{Voort2011a} at all redshifts between $z=0$ and $z=4$, both in \Mstream\ as in general shape. We only note a tendency for the fraction estimates from of \citeauthor{Voort2011a} to assume higher values (on average lower than 10~per~cent) than our counterparts, in particular at $\log\Mvir\gtrsim11.5\,\Msun$. When comparing all the entries in Table~\ref{tab:literature_methods} between these authors and our work, we note that their study is the closest to ours in terms of code, sub-grid physics, and the method employed. It is thus unlikely to be a coincidence that the resulting \fcold\ exhibits the best agreement and underlines the difficulty in comparing various studies together.

We end this section by exploring how \Mshock\ and \Mstream\ evolve with redshift, recalling first their definition. We define \Mshock\ as the transition mass where $\mathcal{S}_1=0.5$ and \Mstream\ as the transition mass where $\mathcal{S}_2 = 0.5$. In our model, \Mshock\ is found to be relatively constant with redshift, while \Mstream\ increases with redshift and the transition between the two regimes occurs smoothly around $z \sim 1.2$. Different authors might adopt varying definitions of \Mshock\ and \Mstream\ with, e.g., \citet{Ocvirk2008} constraining \Mshock\ to the accretion onto the galaxy and \Mstream\ to the accretion onto the halo. This distinction should be borne in mind when interpreting the subsequent figure.

The general form of our critical mass separating cold and hot dominated accretion is consistent with \citet{Dekel2006}: below the transition redshift $z_\tau \sim 1.3$, the critical mass (\Mshock) is relatively constant at $\log(M_c/\Msun)\sim 12$ before transitioning to an evolving critical mass \Mstream, with our inferred $z_\tau$  consistent with the value of $\sim1.4$ found by \citet{Dekel2006}. The evolution of \Mstream, however, shows a relatively large discrepancy with the latter. Importantly, their predictions are based on idealised spherical accretion models and relatively high metallicities assumed for the gas in filaments. Indeed, \citet{Ocvirk2008} show that if they correct for lower metallicities, they can bring the model from \citet{Dekel2006} in agreement with their data, which would consequently reduce the disagreement with our results as well. 

As for the comparison between \citet{Ocvirk2008} and our work, we retrieve the same characteristics seen in Fig.~\ref{fig:model_comparison}. Their \Mstream\ remains relatively low up to $z=3$ ($\Mstream<6\times10^{11}\,\Msun$), after which it manifests a large increase to $z=4$, reaching $\Mstream\approx10^{13}\Msun$. On the other hand, their estimate for \Mshock\ remains nearly stable at the value of at all redshifts. As we pointed out earlier, the differences in code and method render a fair comparison rather difficult. It is nonetheless interesting to note that our extrapolated values for \Mshock\ are quite consistent with their measurements.

Finally, by using observed data from \citet{Lee2015} and \citet{Delvecchio2021} and analysing the stellar mass at which the SFMS flattens \citep[see, e.g.][]{Schreiber2015, Tomczak2016,Popesso2023} and converting it to DM mass, \citet{Daddi2022b} find this transition in the SFMS to be in notable agreement with the \Mstream\ prediction from \citet{Dekel2006}. The dotted line in Fig.~\ref{fig:mcrit_comparison} is fit to the transition mass performed by \citet{Daddi2022b}. Although the error bars on the original data are substantial at high-$z$ (see their Fig.~2, right panel), these results are nonetheless promising for advancing our understanding of the complex relationship between cold accretion in galaxies and their SFRs.

\subsection{Limitations of the model}
Before summarising and concluding this work, we want to address some limitations and caveats of the model presented in this work. As already pointed out, the nature and specific implementation of different feedback models can vary significantly among simulations, rendering precise comparisons difficult. The addition of various feedback mechanisms influences the relative contribution of the two accretion modes and thus the transition mass between cold and hot-dominated accretion. However, \citet{Voort2011a} find that hot accretion fraction onto haloes is insensitive to feedback and metal-line cooling, as opposed to the accretion onto the galaxy. Furthermore, results from \citet{Correa2018a} suggest that the transition critical mass depends mostly on the cooling rate rather than any of the feedback parameters. \footnote{Gaining deeper insight into how SN and AGN feedbacks shape the evolution of gas accretion modes remains essential. To address this aspect, we will update our code for the next generation of our simulations in order to more accurately track the impact of individual feedback mechanisms on each gas particle.}

Cooling times strongly depend on the metallicity content of the gas, with higher metallicities enhancing its ability to cool efficiently. Consequently, metallicity is expected to impact the critical mass at which cold-dominated accretion transitions to hot-dominated accretion \citep[][]{Birnboim2003,Dekel2006}. For instance, \citet{Ocvirk2008} demonstrate that varying assumptions about primordial metallicities in gas filaments can significantly shift the critical mass for this high-$z$ transition. The early enrichment of the IGM by the first stars is thus predicted to play a role in shaping the critical mass.
\citet{Wiersma2009a}, however, show that metal-line cooling becomes significant only when metallicities exceed $Z \sim 0.1\ Z_\odot$. This finding is further supported by \citet{Correa2018a}, who report that higher metallicities correspond to higher critical masses; nevertheless, for metallicities below 10~per~cent of $Z_\odot$, the normalisation of the critical mass remains almost unaffected. In HELLO galaxies, star formation is well resolved up to at least $z \sim 8$, an epoch where the mean stellar metallicity is approximately $0.1\ Z_\odot$ \citep[][]{Wiersma2009b}. Despite this, our current model does not account for the metallicity dependence explicitly and we plan to address this limitation in the future by incorporating the metallicity dependence alongside redshift and mass.

Another potential caveat of our analysis regards the complexity of the model adopted to describe the evolution of the relation between the cold fraction of gas and the viral mass (refer to \S~\ref{sec:fcold_cosmic_evolution}). In light of that, in Appendix~\ref{app:single_vs_double} we addressed this possible caveat, by comparing the double-sigmoid model presented in the main body of the paper to a simpler description, namely a single-sigmoid model.

\section{Summary and conclusion}\label{sec:conclusion}
In this manuscript, we investigated the cold gas accretion onto massive simulated galaxies ($\Mvir \sim [10^{12}-10^{13}]\Msun$) from $z=0$ to beyond the cosmic noon using a combination of galaxies from the NIHAO and HELLO suites. Our primary focus was to understand how the cold fraction of gas around galaxies evolves over cosmic time and how this evolution influences the transition between cold- and hot-dominated accretion modes. We employed as our fiducial method a Lagrangian approach, considering the maximum historical temperature of gas particles to distinguish between cold and hot accretion, allowing us to track accretion modes consistently across time. We ultimately adopted a Bayesian hierarchical formalism to model the continuous evolution in the redshift of \fcold as a function of the viral mass of systems and compared our results with various previous studies in the literature. We summarise our results as follows:

(i) HELLO galaxies at $z=3.6$ and $z=2$ exhibit large inflows of cold (as well as hot) gas from the IGM down to the galaxy. A significant fraction of that gas is fueling the central galaxies through streams of cold gas able to penetrate the halo unshocked. At the same time, the space between these filaments is filled by hot outflows of stellar feedback-induced winds, expanding beyond \Rvir\ and affecting the inflowing gas radial temperature, increasing the fraction of hot gas accreting onto the halo in the vicinity of \Rvir.

(ii) Instantaneous accretion rates are an order of magnitude higher in HELLO simulations compared to NIHAO, with inflows through \Rvir\ of $\sim\!110\,\Msun\,\mathrm{yr}^{-1}$ at $z=2$ and $\sim\!240\,\Msun\,\mathrm{yr}^{-1}$ at $z=3.6$. However, outflows are roughly matching the inflows at $z=2$, resulting in low net accretion rates of $\sim\!20\,\Msun\,\mathrm{yr}^{-1}$ , but are subdominant at $z=3.6$, giving net rates of $\sim\!130\,\Msun\,\mathrm{yr}^{-1}$.

(iii) Cold accretions rates within the halo are of the order $10^2$, few $\times 10$, and order 1 $\Msun,\mathrm{yr}^{-1}$ at $z=3.6$, $z=2$, and $z=0$, respectively.

(iv) At the respective final redshifts, HELLO systems show a significantly higher average \fcold for particles entering the galaxy compared to those from NIHAO. Specifically, HELLOz3.6 and HELLOz2.0 exhibit cold gas fractions of approximately 90~per~cent and 70~per~cent, respectively, while NIHAO reaches only about 20~per~cent. This suggests a strong dependence on redshift for galactic accretion. In contrast, the redshift dependence for accretion onto the halo is weaker, with cold gas fractions around 40~per~cent, 25~percent, and 20~per~cent for HELLOz3.6, HELLOz2.0, and NIHAO, respectively.

(v) Modelling the continuous evolution of \fcold accreting onto the halo with redshift and mass, our results align well with previous studies at $z=0$, but reveal discrepancies at higher redshifts due to differences in the employed method to obtain \fcold\ and as well as the implemented sub-grid physics.

(vi) We find an almost constant critical mass \Mshock\ around $\log(\Mshock / \Msun) \sim 11.8$ up to a transition redshift $z \sim 1.3$ and an evolving $\Mstream \propto \log(1+z)^{1.7}$ afterwards.

(vii) The inferred critical mass for the transition between cold and hot accretion is consistent with theoretical predictions by \citet{Dekel2006} in the low-$z$ ($z\lesssim1.3$) regime but shows discrepancies at higher redshifts, likely due to the idealised nature of the predictions \citep[see also][]{Ocvirk2008}.
\\

The results from our analysis extend to previous work and confirm the complex relation between redshift and mass in determining the dominant accretion mode in galaxies. Our high-$z$ galaxies exhibit a significant amount of cold accretion, driven primarily by cold streams in DM filaments radially intersecting at nodes in the centre of which galaxies experience a continuous flow of fresh cold gas able to sustain their high SFRs. In contrast, galaxies at $z=0$ exhibit vanishing cold as well as hot accretion rates and are embedded in a stable and extended hot halo reaching beyond their virial radius.

Our Bayesian hierarchical model captures the evolution of \fcold\ between the high- and low-$z$ regimes and predicts a relatively constant critical mass for cold-to-hot transition (\Mshock) up to $z \sim 1.3$, remarkably consistent with the prediction of $z \sim 1.4$ from \citet{Dekel2006}. When comparing our results with other simulations \citep[][]{Ocvirk2008, Voort2011a, Correa2018a}, we find good agreement at low redshift, but some divergence at higher redshift due to differences in several aspects, such as feedback models, temperature criteria, and accretion methods, highlighting the complex interplay between positive and negative feedback mechanisms at high redshift.

While our results show some discrepancies with the cold fraction results inferred from observational data by \citet{Daddi2022b}\footnote{It is worth noticing that several assumptions are needed to extract $M_c$ from observations since this is not a directly observable quantity.}, their findings are nonetheless encouraging and accentuate the importance of further constraining the methods in observations and simulations alike to better understand the intricate relationship between gas accretion and SFRs. Overall, our findings provide deeper insight into the cosmic evolution of gas accretion modes and offer a robust framework for understanding how cold accretion contributes to galaxy growth across different epochs.


\section*{Acknowledgements}
We sincerely thank the referee Y. Birnboim
for the feedback that considerably contributed to improving the quality of this manuscript.
This material is based upon work supported by Tamkeen under the NYU Abu Dhabi Research Institute grant CASS.
The authors gratefully acknowledge the support of 
the High Performance Computing resources at New York University Abu Dhabi, where most of the simulations were performed.

\section*{Data availability}
The data underlying this article will be shared on reasonable
request to the corresponding author.



\bibliographystyle{mnras}
\bibliography{ref} 




\appendix

\section{Cold fractions from various methods}\label{app:cold_frac_evol}

\begin{figure*}
\centering
\includegraphics[width=\linewidth]{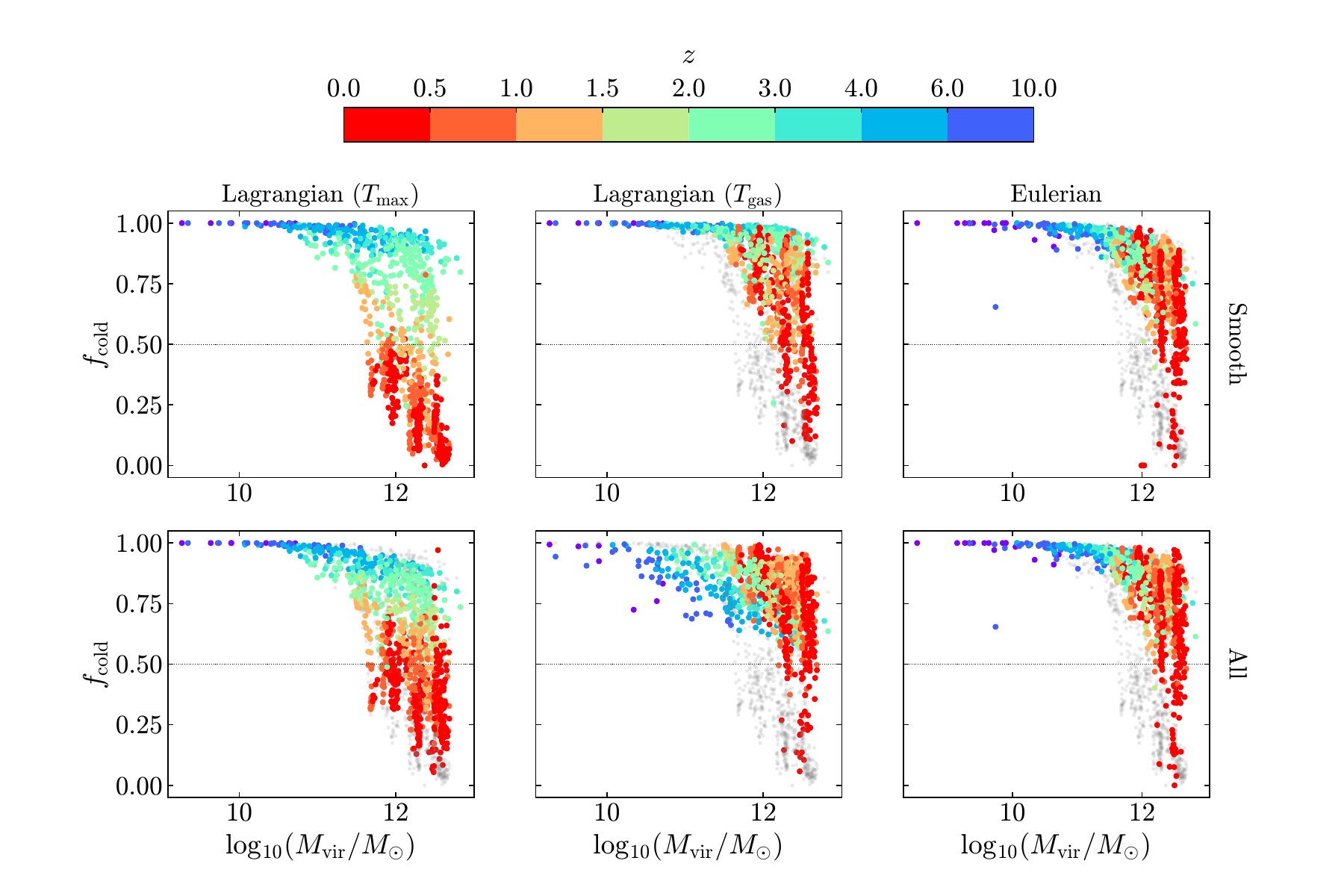}
\caption{Cold fraction evolutionary tracks from particles crossing the galaxy using smooth particles (top) and all particles (bottom). Each column represents a different method to compute \fcold: maximum historical gas temperature across all previous snapshots (\Tmax; left), gas temperature right before entering the galaxy ($T_{\mathrm{gas}}$; middle), and temperature of inflows through a shell around \Rgal (Eulerian; right) Our fiducial method (upper left) is shown as a grey background in each subsequent panel.}
\label{fig:cold_frac_evolution_gal_comp}
\end{figure*}

\begin{figure*}
\centering
\includegraphics[width=\linewidth]{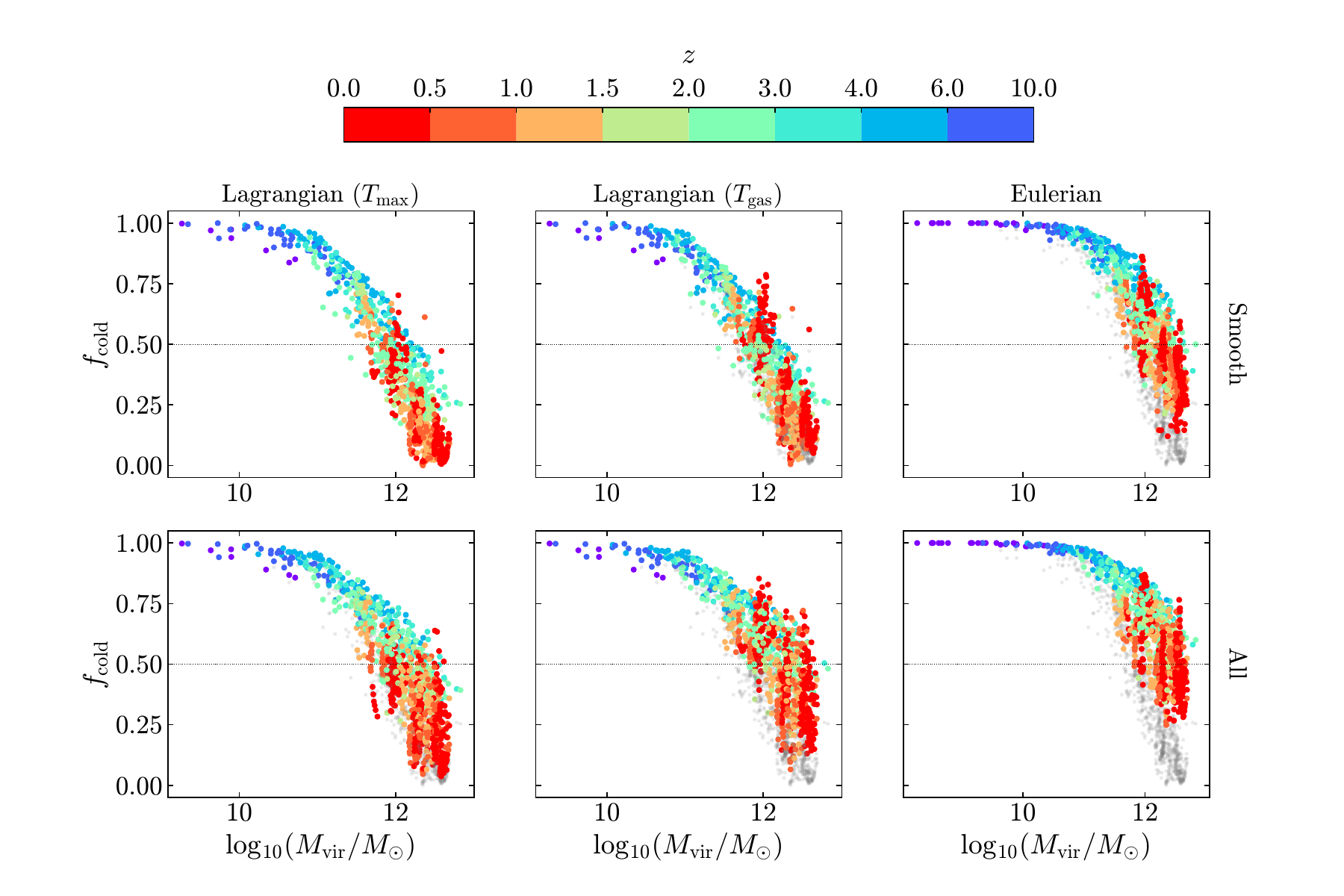}
\caption{Cold fraction evolutionary tracks from particles crossing the halo using smooth particles (top) and all particles (bottom). Each column represents a different method to compute \fcold: maximum historical gas temperature across all snapshots (\Tmax; left), gas temperature right before entering the galaxy ($T_{\mathrm{gas}}$; middle), and temperature of inflows through a shell around \Rgal (Eulerian; right) Our fiducial method (upper left) is shown as a grey background in each subsequent panel.}
\label{fig:cold_frac_evolution_halo_comp}
\end{figure*}
In Fig.~\ref{fig:cold_frac_evolution}, we presented the distributions of \fcold as a function of the viral mass for smooth particles crossing \Rgal\ and \Rvir, highlighting the three simulation sets. It is worth checking whether the above-mentioned distributions vary when adopting different criteria to define \fcold as well as using all particles.
The accretion onto the galaxy is shown in Fig.~\ref{fig:cold_frac_evolution_gal_comp}, while the accretion onto the halo is plotted in Fig.~\ref{fig:cold_frac_evolution_halo_comp}. For easier comparison, in each plot, the upper-left panels collect the data points shown in Fig.~\ref{fig:cold_frac_evolution}.
We stress here that the focus of our work is not to quantify how \fcold\ varies with definition and even in the case of accretion through \Rvir\ differences are visible. Compared to the accretion onto \Rgal\ though, the figures suggest \textit{less} of an impact. For these reasons, in the main part of this work, we focus on \fcold\ at \Rvir, determined from our fiducial method (smooth gas, Lagrangian, and \Tmax\ criterion).

\section{The Bayesian hierarchical framework}\label{app:bayesian}
Our strategy to trace the evolution of \fcold through cosmic time is to exploit a Bayesian 
hierarchical approach. In this context, each object can be represented by a set of three physical properties:
\begin{itemize}
    \item the virial mass \Mvir;
    \item the cold gas fraction \fcold;
    \item the redshift $z$.
\end{itemize}
We refer to these three parameters as $\Params=\{\Mvir,\fcold,z\}$. We can consider the three properties as drawn from corresponding probability distributions which can be described by a set of global hyperparameters $\Hyper$, so that $\Prob(\Params)=\Prob(\Params|\Hyper)$.

From the Bayes' theorem, the posterior probability distribution of $\Hyper$ given data $\Data$ is
\begin{equation}
    \Prob(\Hyper|\Data)=\frac{\Prob(\Data|\Hyper)\,\Prob(\Hyper)}{\Prob(\Data)},
    \label{eq:bayes}
\end{equation}
with $\Prob(\Data)\neq0$.
In Eq.~\ref{eq:bayes}, $\Prob(\Data|\Hyper)$ is the \emph{likelihood} function of data given a model, $\Prob(\Hyper)$ is the \emph{prior probability distribution}, and $\Prob(\Data)$ is the \emph{Bayesian evidence}.

Since our data $\Data=\{M_\mathrm{vir}^\mathrm{sim},f_\mathrm{cold}^\mathrm{sim},z^\mathrm{sim}\}$ are derived from simulations, no a priori uncertainties are associated with them.
This implies that when we evaluate the likelihood term in Eq.~\ref{eq:bayes} for the $i$-th object, the $\Prob(\Data_i|\Theta_i)$ term reduces to a delta function
\begin{equation}
\begin{aligned}
    \Prob(\Data_i|\Hyper) = &\int_{\Params_i \in \Params} \, \diff\Params_i\,\Prob(\Data_i|\Params_i)\,\Prob(\Params_i|\Hyper)=\\
    =&\int_{\Params_i \in \Params}\diff\Params_i\,\delta(\Data_i-\Params_i)\Prob(\Params_i|\Hyper).
    \label{eq:likelihood}
\end{aligned}
\end{equation}

For the relation between \fcold and \Mvir, we assume the probability distribution function $\Prob(\Params|\Hyper)\equiv\Prob(\fcold,\Mvir,z|\Hyper)$ follows a normal distribution
\begin{equation}
    \Prob(\Params|\Hyper) \sim \mathcal{N}(f_\mathrm{cold},\Mvir, z,|\Hyper) = \frac{1}{\sqrt{2\pi\sigma_{\fcold}^2}}
                                      \exp{\left(-\frac{(\fcold-\mu_{\fcold})^2}{2\sigma_{\fcold}^2}\right)},
    \label{eq:normal}
\end{equation}
with $\mu_{\fcold}$ the expected cold fraction from Eq.~\ref{eq:model}.
The standard deviation term in Eq.~\ref{eq:normal} represents the intrinsic scatter of the relation, which we assume evolves with redshift as
\begin{equation}\label{eq:intrinsic_scatter}
    \sigma_{\fcold} = \sigma_{0} + \sigma_{z}\log(1+z),
\end{equation}
with hyperparameters $\sigma_0$ and $\sigma_z$.

In Table~\ref{tab:hyperparameters}, we list all the hyperparameters used in our model, providing a brief description and the priors adopted. Note that we empirically found that setting $\kappa_\tau = -10$ provides a reasonable value, so we have chosen to fix it as a constant. 

\begin{table*}
\begin{center}
    \begin{tabular}{ccc} 
 \hline
 Hyperparameter & Description & Prior\\
 \hline
 $\kappa_{0,1}$   & Normalisation of the steepness in Eq.~\ref{eq:steepness} for the low-$z$  $\mathcal{S}_1$ sigmoid      & $\mathcal{U}$[-6,  0] \\ 
 $\kappa_{z,1}$   & Slope of the steepness in Eq.~\ref{eq:steepness}         for the low-$z$  $\mathcal{S}_1$ sigmoid      & $\mathcal{U}$[-10, 10] \\ 
 $\log M_{0,1}$        & Normalisation of the inflexion in Eq.~\ref{eq:inflexion} for the low-$z$  $\mathcal{S}_1$ sigmoid      & $\mathcal{U}$[10,12.5] \\ 
 $\log M_{z,1}$        & Slope of the inflexion in Eq.~\ref{eq:inflexion}         for the low-$z$  $\mathcal{S}_1$ sigmoid      & $\mathcal{U}$[-5,5] \\ 
 $\kappa_{0,2}$   & Normalisation of the steepness in Eq.~\ref{eq:steepness} for the high-$z$ $\mathcal{S}_2$ sigmoid      & $\mathcal{U}$[-6,  0] \\ 
 $\kappa_{z,2}$   & Slope of the steepness in Eq.~\ref{eq:steepness}         for the high-$z$ $\mathcal{S}_2$ sigmoid      & $\mathcal{U}$[-10, 10] \\ 
 $\log M_{0,2}$        & Normalisation of the inflexion in Eq.~\ref{eq:inflexion} for the high-$z$ $\mathcal{S}_2$ sigmoid      & $\mathcal{U}$[10,12.5] \\ 
 $\log M_{z,2}$        & Slope of the inflexion in Eq.~\ref{eq:inflexion}         for the high-$z$ $\mathcal{S}_2$ sigmoid      & $\mathcal{U}$[-5,5] \\ 
 $\kappa_\tau$         & Steepness in Eq.~\ref{eq:steepness}                      for the transition $\mathcal{S}_\tau$ sigmoid & -10 \\ 
 $z_\tau$         & Inflexion in Eq.~\ref{eq:inflexion}                      for the transition $\mathcal{S}_\tau$ sigmoid & $\mathcal{U}$[0.0,2.2] \\ 
 $\sigma_0$       & Normalisation of the intrinsic scatter of the $\fcold{-}\Mvir$ relation                        & $\mathcal{U}$[0,0.5] \\ 
 $\sigma_z$       & Normalisation of the intrinsic scatter of the $\fcold{-}\Mvir$ relation                        & $\mathcal{U}$[-1,1] \\ 
\hline
\end{tabular}
\caption{Hyperparameters of the model defined in this section.
         Column 1: name of the hyperparameter.
         Column 2: description of the hyperparameter.
         Column 3: range of the uniform prior.}
\label{tab:hyperparameters}
\end{center}
\end{table*}

\section{Single-sigmoid model versus double-sigmoid model}\label{app:single_vs_double}
\begin{figure*}
\centering
\includegraphics[width=\linewidth]{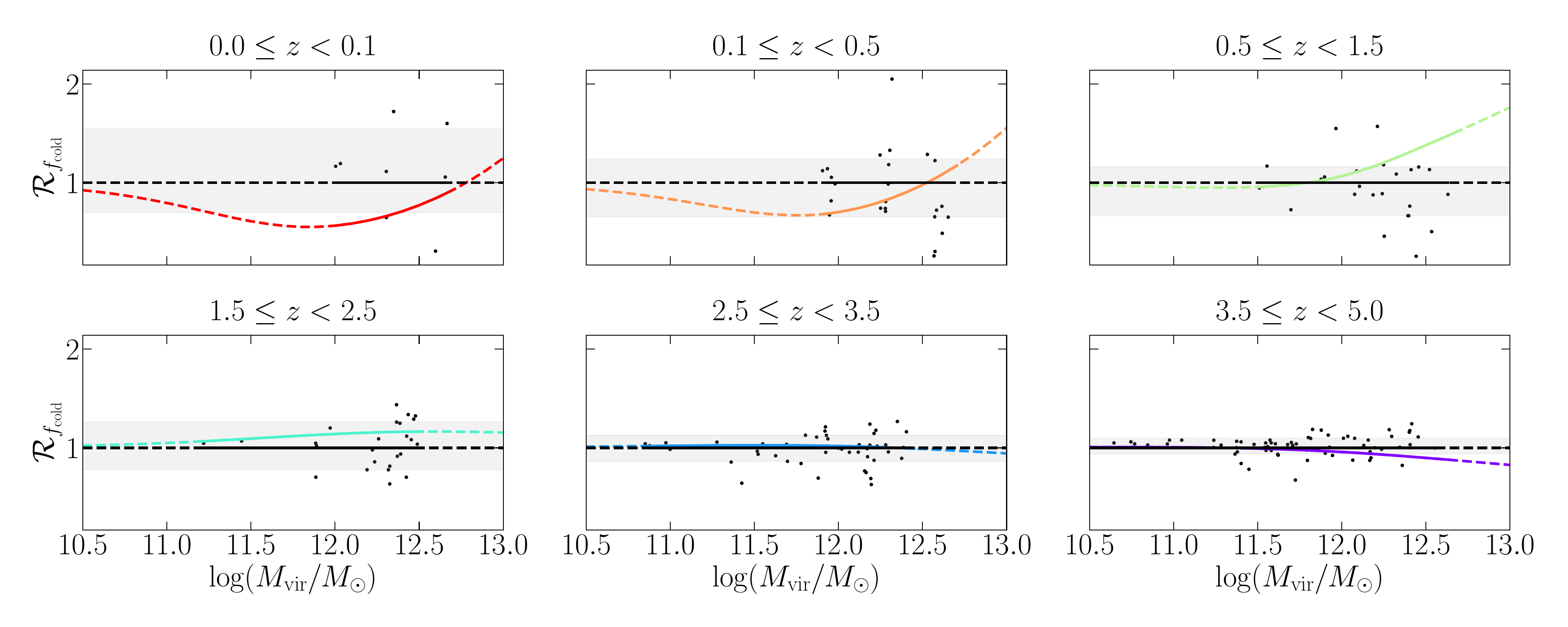}
\caption{Ratio between the cold fraction inferred from the single-sigmoid model $\mathcal{M}_s$ and the double-sigmoid model $\mathcal{M}_d$ (colored curves) as a function of the virial mass in six redshift bins. The black points mark the ratio between the true values of the cold fraction from the simulations and the values from $\mathcal{M}_d$, while the grey-shaded regions enclose the 68 per cent of data in each redshift bin. The black curves represent a ratio equal to 1.}
\label{fig:residual}
\end{figure*}

In \S~\ref{sec:fcold_cosmic_evolution}, we proposed a model to describe the evolution of the relation between the fraction of cold gas and the virial mass that takes into account two different regimes for the critical mass between cold- and hot-dominated accretions  \citep[][]{Dekel2006}. The double-sigmoid model in Eq. \ref{eq:model} might result in a too-flexible model. In light of that, we tested a simpler case adopting a single sigmoid, namely $f_\mathrm{cold}=\mathcal{S}_1$, with $\mathcal{S}_1$ as in Eq.\ref{eq:sigmoid}. In Fig. \ref{fig:residual} we show the ratio $\mathcal{R}_{f_\mathrm{cold}}$ between the fraction from the single-sigmoid model $\mathcal{M}_s$ and from the double-sigmoid model $\mathcal{M}_d$ (the latter presented in the main body of the paper) as a function of the virial mass in six redshift bins. The poor statistics of data at $z\lesssim2$ do not allow the single-sigmoid model to constrain well the evolution of the relation over the entire redshift range explored. Indeed, while at $z\gtrsim2$ the two models are in reasonable agreement with each other, at $z\lesssim2$ the single-sigmoid model underestimates the relation.
Moreover, as reported in \S\ref{ssec:results}, the posterior distributions of the hyperparameters of the models are sampled through the nested sampling technique which allows also us to compute the Bayesian evidence $\mathcal{Z}$ for each model. By computing the logarithm of the Bayes factor of the two models, $\ln\mathcal{B}=\ln\mathcal{Z}_s-\ln\mathcal{Z}_d$, where $\ln\mathcal{Z}_s\simeq525$ and $\ln\mathcal{Z}_d\simeq541$ are related to the single and double-sigmoid models, respectively, we obtain $\ln\mathcal{B}\simeq-16$. Comparing the value of the Bayes factor with the empirical  Jeffreys' scale \citep{Jeffreys1961ToP}, the result lends decisive support to model $\mathcal{M}_d$.

\section{Obtaining independent data}\label{app:indep_data}
In \S~\ref{sec:fcold_cosmic_evolution} we trace the evolution of the \fcold{-}\Mvir\ relation using a subset of data drawn from our simulation sets. As mentioned in \S~\ref{ssec:independent_data}, we extract a subsample of data points from the original samples because of the necessity of including objects that can be considered independent in redshift. To do that, we can consider the evolutionary track of each galaxy as a time series. A \emph{time series} can be summarised as a collection of observations measured over some (fixed) sampling intervals. For each of our galaxies, we have a sample of cold fractions calculated at each snapshot and separated by fixed time intervals of $\sim\!200$ Myr. Each of these cold fraction evolutionary tracks thus forms a time series, whose observations can be assumed correlated: the cold fraction at some cosmic time $t$ is likely correlated to the cold fraction at $t-1$ and so on.

Statistically, a time series \textit{process} is a sequence of random variables whose realisations constitute the time series. Formally, a time series process is a set of random variables $\left\{X_{t}, t \in T\right\}$, where $T$ is the set of times at which the process was, will or can be observed. Each random variable $X_t$ is assumed to be distributed according to some univariate distribution function $F_t$. The (observed) time series is then the realisation of the random vector $X=(X_1,X_2,...,X_n)$, denoted $x=(x_1,x_2,...,x_n)$.

Our goal is to select for each galaxy data points that can be statistically considered uncorrelated and thus interpreted as two independent observations. For this, we need to determine the serial correlation within each track. The \textit{autocorrelation} between two random variables $X_{t+k}$ and $X_t$ is defined as
\begin{equation}
    \mathrm{Cor}(X_{t+k},X_t) = \frac{\mathrm{Cov}(X_{t+k},X_t)}{\sqrt{\mathrm{Var}(X_{t+k})\,\mathrm{Var}(X_t))}},
\end{equation}
where Cov, Var, and $k$ are the covariance, the variance, and the lag, respectively.

For the autocorrelation to be interpreted adequately, the time series needs to fulfil the requirement of \textit{stationarity}. In simple terms, its mean and variance should be constant over time. For a stationary time series whose moments are constant over time, the autocorrelation only depends on the lag $k$ and we can drop the index $t$
\begin{equation}
    \rho(k) \equiv \mathrm{Cor}(X_{t+k}, X_t).
\end{equation}

In practice, time series usually display trend and/or seasonal effects which violate stationarity. In such cases, one can use decomposition models such as, e.g., the simple additive decomposition model
\begin{equation}
    X_t = m_t + s_t + R_t,
\end{equation}
where the time series is decomposed into its trend $m_t$, its seasonality $s_t$ and its remainder $R_t$. Ideally, such a decomposition should lead to a stationary remainder that can then be used for statistical inference. Since our \fcold tracks decrease between one and zero over time, our time series have a clear trend. For simplicity, we assume no seasonality and model the underlying trend as a sigmoid $\mathcal{S}$, i.e.
\begin{equation}
    f_{\mathrm{cold},t} = \mathcal{S}_t + R_t.
\end{equation}

We first look at each individual track and remove the few chaotic ones that are inconsistent with a (rough) sigmoidal evolution. In total, we remove five galaxies from the NIHAO sample, leaving us with eight galaxies plus both HELLO samples. We then apply the methods of least squares to fit a sigmoid to each track and compute $\rho(k)$ on the residuals. To determine the lags that can be statistically considered independent we compute the 95~per~cent confidence bands $\pm 1.96 / \sqrt{N}$, where $N$ is the number of observations in the time series. Autocorrelation coefficients that fall within this interval can be considered different from zero only by chance.

\begin{figure*}
\centering
\includegraphics[width=\linewidth]{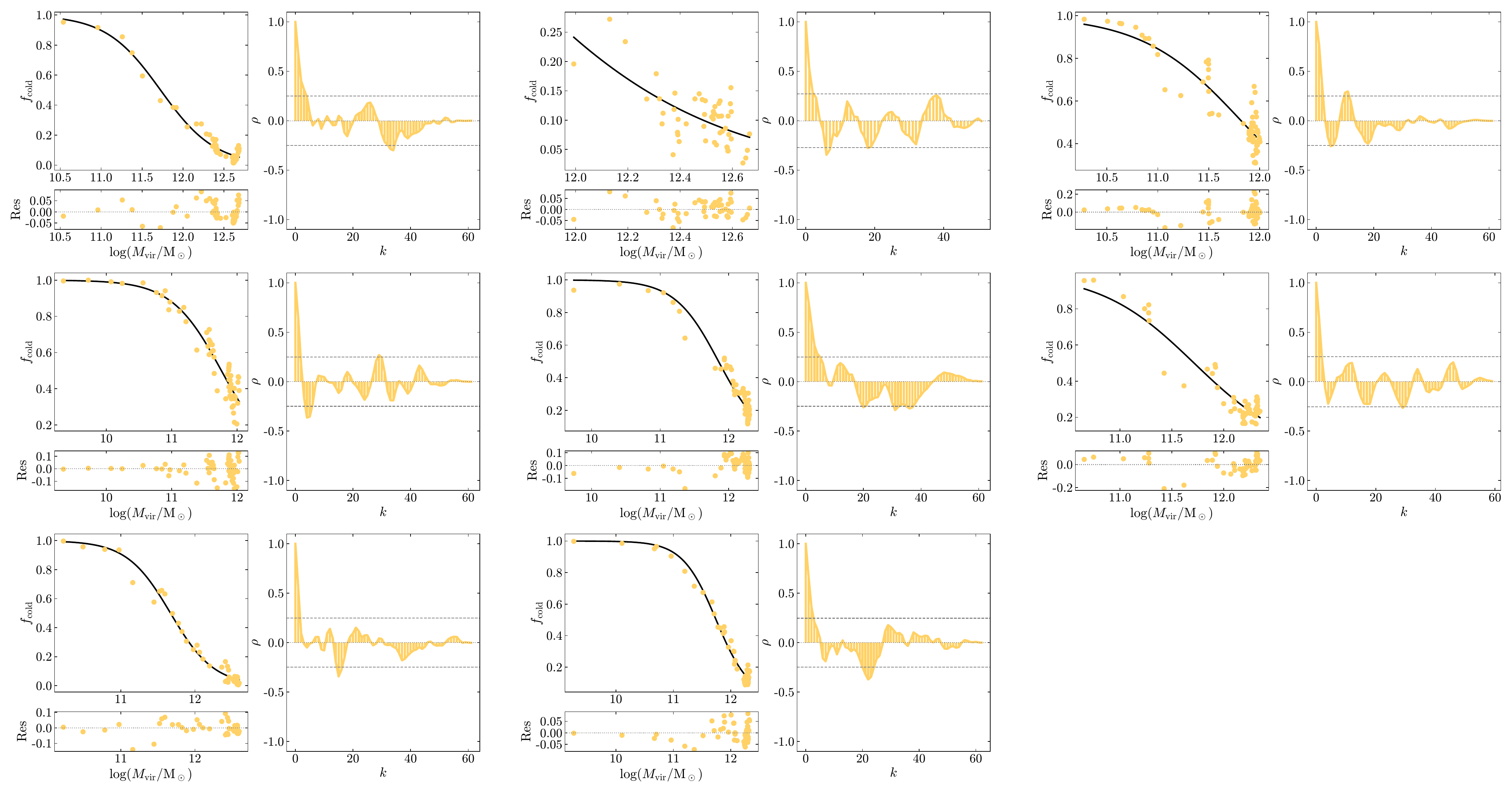}
\caption{Autocorrelations in \fcold for NIHAO galaxies after detrending the time series. Each galaxy is shown in a panel containing three subpanels: the track with the sigmoid fit (upper left), the residual (remainder) after removing the trend (lower left), and the correlogram of the ACF as a function of the lag $k$ (right). The grey dashed lines in each correlogram represent the 95~per~cent confidence interval.}
\label{fig:nihao_acf}
\end{figure*}

\begin{figure*}
\centering
\includegraphics[width=\linewidth]{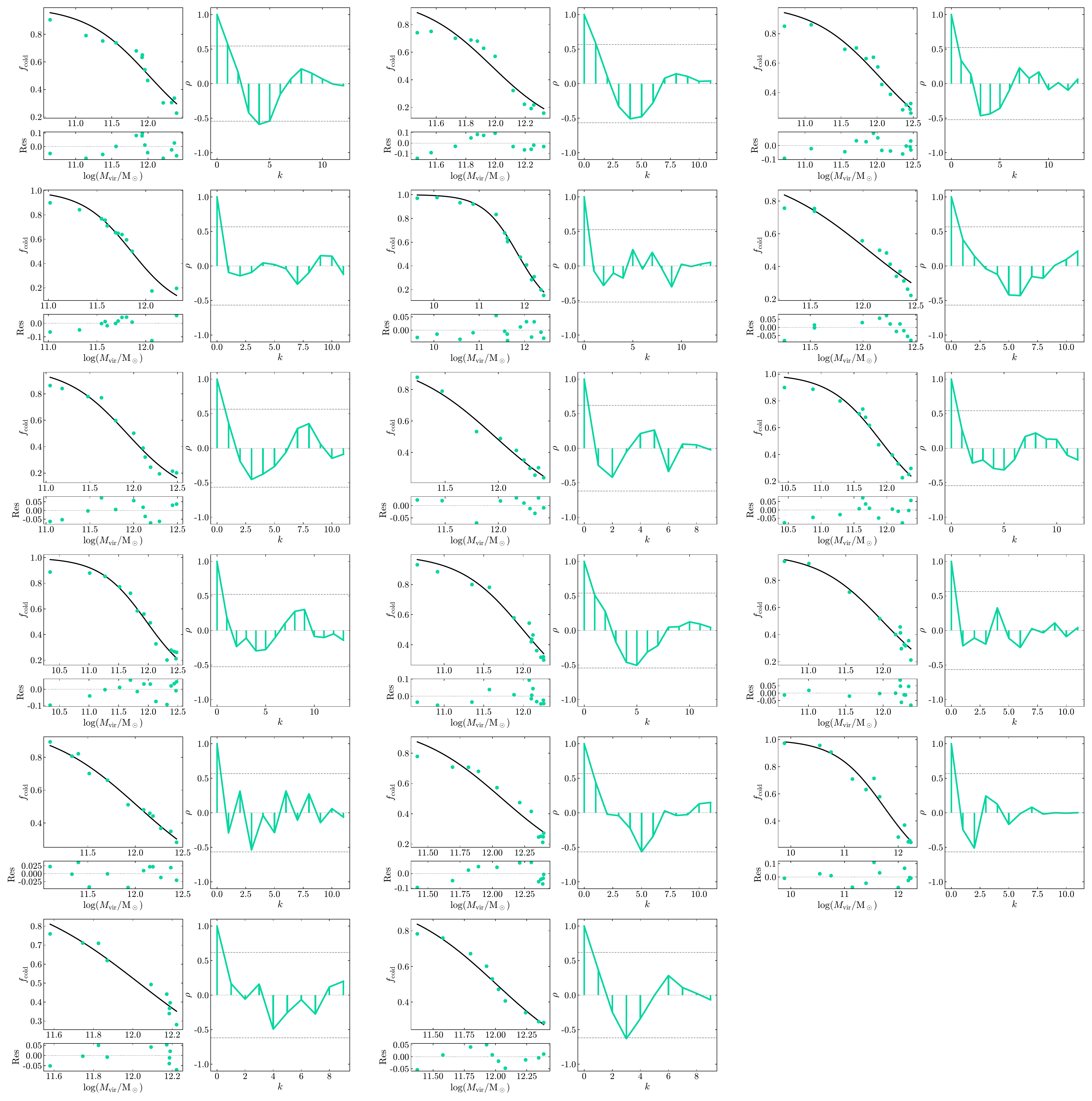}
\caption{Same as in Fig.~\ref{fig:nihao_acf}, but for HELLOz2.0 galaxies.}
\label{fig:hello2_acf}
\end{figure*}

\begin{figure*}
\centering
\includegraphics[width=\linewidth]{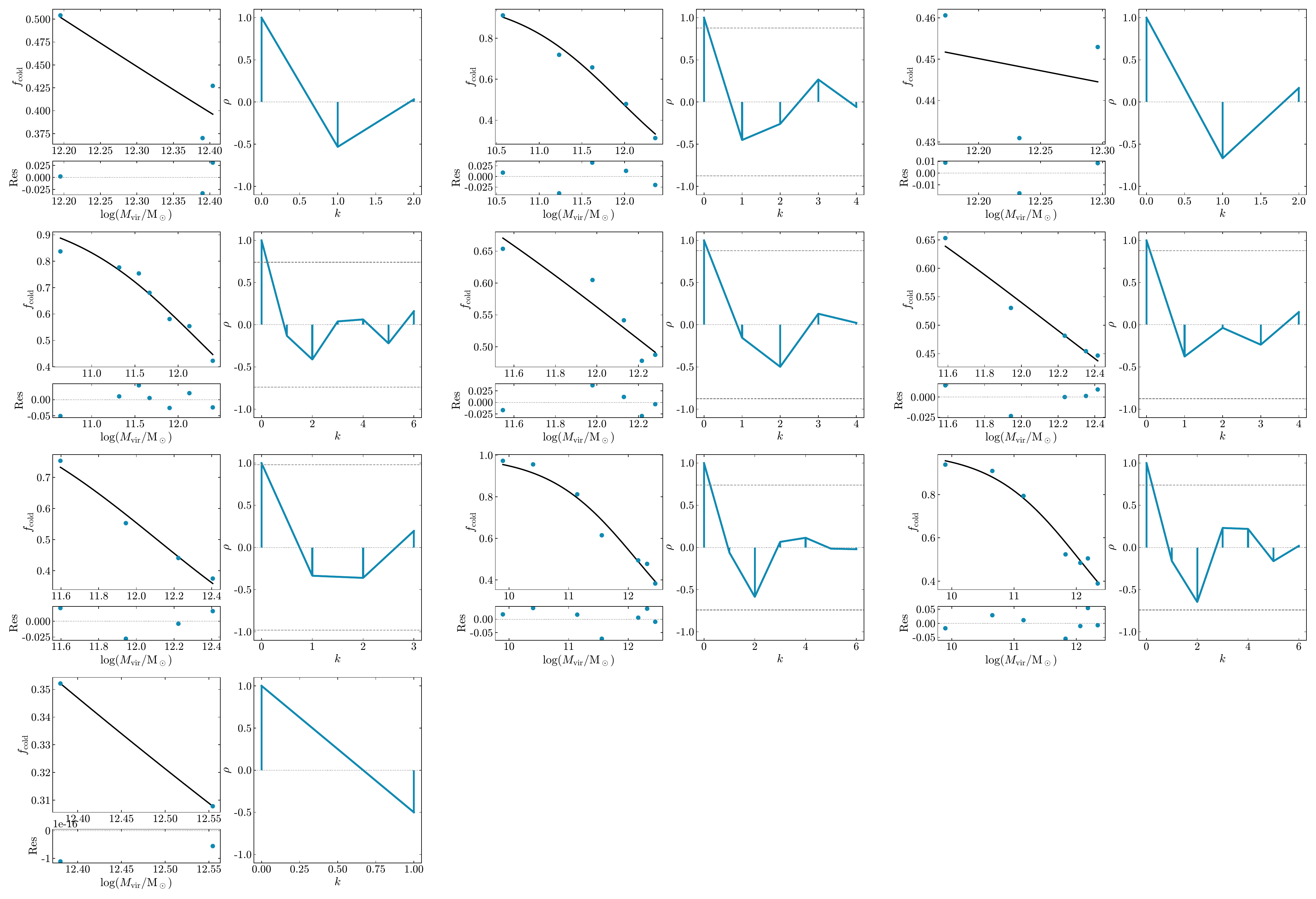}
\caption{Same as in Fig.~\ref{fig:nihao_acf}, but for HELLOz3.6 galaxies.}
\label{fig:hello3_acf}
\end{figure*}

The results for each sample can be seen in Figs.~\ref{fig:nihao_acf}, \ref{fig:hello2_acf}, and \ref{fig:hello3_acf}. It is important to note that the low number of points per track in HELLO galaxies naturally leads to wider confidence intervals, thereby diminishing the statistical significance of these results. Despite this, all correlograms display a consistent pattern, with $\rho$ rapidly decaying towards zero and remaining relatively low. The fact that all coefficients remain small at large lags indicates that the trend has been successfully removed from each track. Some correlograms exhibit a potential seasonal component, evidenced by their cyclical behaviour. However, since these patterns remain within the confidence intervals, we do not find it necessary to revise our assumption of no seasonality.

The final step is to determine the lag at which observations can be considered independent. On one hand, we aim to be conservative, avoiding the use of consecutive snapshots even though some correlograms already show minimal correlations at $k=1$. On the other hand, maximizing redshift coverage is a necessity. Additionally, we require a consistent lag across all galaxies within a sample. Considering these criteria, we find that $k=7$ for NIHAO, and $k=3$ for HELLO provide an acceptable balance between independence and data coverage.

\section{Evolution of the model parameters}
\begin{figure}
\centering
\includegraphics[width=\linewidth]{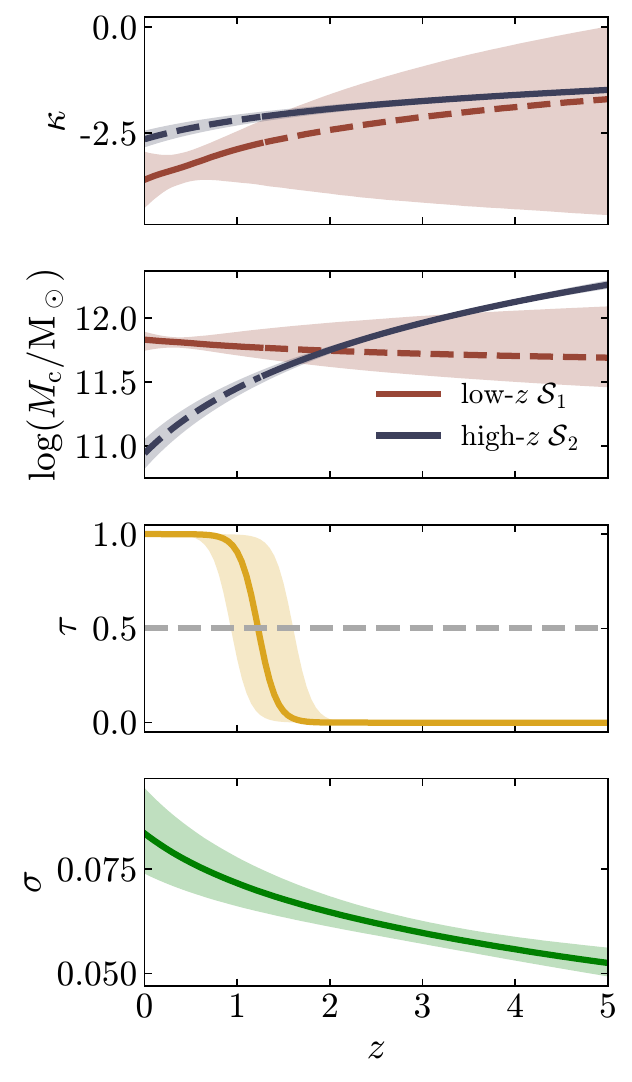}
\caption{Redshift evolution of each parameter. The top panel shows the evolution of the steepness $\kappa$ (Eq.~\ref{eq:steepness}) for both the low-$z$ regime (red) and the high-$z$ regime (blue) sigmoids. The shaded regions cover the respective 16th-84th percentile interval. The second panel is similar but for the evolution of the critical mass according to Eq.~\ref{eq:inflexion}. The third panel shows the evolution of $\tau$, the coefficient for the transition between the two regimes. Finally, the lower panel shows the evolution of the intrinsic scatter as per Eq.~\ref{eq:intrinsic_scatter}.}
\label{fig:hparams_evo}
\end{figure}

In \S~\ref{ssec:results}, we provide an approximated description of how \fcold evolves with redshift as a function of \Mvir. Reminding the reader that to model this evolution we adopt a hybrid model consisting of two sigmoids describing the behavior of the fraction of cold gas below and above a redshift transition (refer to Eq.~\ref{eq:model}), in Fig.~\ref{fig:hparams_evo} we show how the steepness and characteristic masses of the two sigmoids evolve between z=0 and z=5, as well as the transition between the two regimes in redshift and the intrinsic scatter of the data distribution.

\bsp	
\label{lastpage}
\end{document}